\documentclass[conference,a4paper]{IEEEtran}
\usepackage{mdwmath}
\usepackage{mdwtab}
\usepackage{graphicx}
\usepackage{amsmath}
\usepackage{latexsym}
\usepackage{amssymb}
\usepackage{comment}
\usepackage{cite}
\usepackage{url}
\usepackage[T3,T1]{fontenc}
\DeclareSymbolFont{tipa}{T3}{cmr}{m}{n}
\DeclareMathAccent{\invbreve}{\mathalpha}{tipa}{16}

\global\long\def\L{\mathsf{L}}
\global\long\def\D{\mathsf{D}}
\global\long\def\Sgen{\mathcal{S}_\mathsf{gen}}
\global\long\def\Kgen{\mathcal{K}_\mathsf{gen}}
\global\long\def\A{\mathcal{A}}

\global\long\def\SgenOne{\mathcal{S}_\mathsf{gen,1}}
\global\long\def\SgenTwo{\mathcal{S}_\mathsf{gen,2}}

\newcommand{\calVarX}{{\cal X}}

\newcommand{\E}{\mbox{\bf E}}

\newcommand{\Dist}[1]{{#1}}

\begin{document}
%
\title{
Information Theoretic Security for Broadcasting of Two Encrypted Sources 
under Side-Channel Attacks
}
\author{%
	\IEEEauthorblockN{Bagus Santoso and Yasutada Oohama}
	\IEEEauthorblockA{University of Electro-Communications, Tokyo, Japan\\ 
	Email: \url{{santoso.bagus,oohama}@uec.ac.jp}}
%
}
\maketitle

\begin{abstract}
We consider the secure communication problem for broadcasting of two 
encrypted sources. The sender wishes to broadcast two secret messages 
via two common key cryptosystems. We assume that the adversary can use 
the side-channel, where the side information on common keys can be 
obtained via the rate constraint noiseless channel.      To solve this 
problem we formulate the post encryption coding system. On the 
information leakage on two secrete messages to the adversary, we provide 
an explicit sufficient condition to attain the exponential decay of this 
quantity for large block lengths of encrypted sources.
\end{abstract}
%
\IEEEpeerreviewmaketitle

\newcommand{\qed}{\hfill$\square$}
\newcommand{\suchthat}{\mbox{~s.t.~}}
\newcommand{\markov}{\leftrightarrow}

\newcommand{\argmax}{\mathop{\rm argmax}\limits}
\newcommand{\argmin}{\mathop{\rm argmin}\limits}

\newcommand{\ExP}{\rm e}

\newcommand{\Cls}{class NL}
\newcommand{\vSpa}{\vspace{0.3mm}}
\newcommand{\Prmt}{\zeta}
\newcommand{\pj}{\omega_n}

\newfont{\bg}{cmr10 scaled \magstep4}
\newcommand{\bigzerol}{\smash{\hbox{\bg 0}}}
\newcommand{\bigzerou}{\smash{\lower1.7ex\hbox{\bg 0}}}
\newcommand{\nbn}{\frac{1}{n}}
\newcommand{\ra}{\rightarrow}
\newcommand{\la}{\leftarrow}
\newcommand{\ldo}{\ldots}
\newcommand{\typi}{A_{\epsilon }^{n}}
\newcommand{\bx}{\hspace*{\fill}$\Box$}
\newcommand{\pa}{\vert}
\newcommand{\ignore}[1]{}


\newtheorem{proposition}{Proposition}
\newtheorem{definition}{Definition}
\newtheorem{theorem}{Theorem}
\newtheorem{lemma}{Lemma}
\newtheorem{corollary}{Corollary}
\newtheorem{remark}{Remark}
\newtheorem{property}{Property}
\newtheorem{pr}{Property}
\newtheorem{Ex}{Example}

\newcommand{\defeq}{:=}

\newcommand{\Qed}{\hbox{\rule[-2pt]{3pt}{6pt}}}
\newcommand{\beq}{\begin{equation}}
\newcommand{\eeq}{\end{equation}}
\newcommand{\beqa}{\begin{eqnarray}}
\newcommand{\eeqa}{\end{eqnarray}}
\newcommand{\beqno}{\begin{eqnarray*}}
\newcommand{\eeqno}{\end{eqnarray*}}
\newcommand{\ba}{\begin{array}}
\newcommand{\ea}{\end{array}}

\newcommand{\vc}[1]{\mbox{\boldmath $#1$}}
\newcommand{\lvc}[1]{\mbox{\scriptsize \boldmath $#1$}}
\newcommand{\svc}[1]{\mbox{\scriptsize\boldmath $#1$}}

\newcommand{\wh}{\widehat}
\newcommand{\wt}{\widetilde}
\newcommand{\ts}{\textstyle}
\newcommand{\ds}{\displaystyle}
\newcommand{\scs}{\scriptstyle}
\newcommand{\vep}{\varepsilon}
\newcommand{\rhp}{\rightharpoonup}
\newcommand{\cl}{\circ\!\!\!\!\!-}
\newcommand{\bcs}{\dot{\,}.\dot{\,}}
\newcommand{\eqv}{\Leftrightarrow}
\newcommand{\leqv}{\Longleftrightarrow}

\newcommand{\irr}[1]{{\color[named]{Red}#1\normalcolor}}

\newcommand{\hugel}{{\arraycolsep 0mm
                    \left\{\ba{l}{\,}\\{\,}\ea\right.\!\!}}
\newcommand{\Hugel}{{\arraycolsep 0mm
                    \left\{\ba{l}{\,}\\{\,}\\{\,}\ea\right.\!\!}}
\newcommand{\HUgel}{{\arraycolsep 0mm
                    \left\{\ba{l}{\,}\\{\,}\\{\,}\vspace{-1mm}
                    \\{\,}\ea\right.\!\!}}
\newcommand{\huger}{{\arraycolsep 0mm
                    \left.\ba{l}{\,}\\{\,}\ea\!\!\right\}}}
\newcommand{\Huger}{{\arraycolsep 0mm
                    \left.\ba{l}{\,}\\{\,}\\{\,}\ea\!\!\right\}}}
\newcommand{\HUger}{{\arraycolsep 0mm
                    \left.\ba{l}{\,}\\{\,}\\{\,}\vspace{-1mm}
                    \\{\,}\ea\!\!\right\}}}

\newcommand{\hugebl}{{\arraycolsep 0mm
                    \left[\ba{l}{\,}\\{\,}\ea\right.\!\!}}
\newcommand{\Hugebl}{{\arraycolsep 0mm
                    \left[\ba{l}{\,}\\{\,}\\{\,}\ea\right.\!\!}}
\newcommand{\HUgebl}{{\arraycolsep 0mm
                    \left[\ba{l}{\,}\\{\,}\\{\,}\vspace{-1mm}
                    \\{\,}\ea\right.\!\!}}
\newcommand{\hugebr}{{\arraycolsep 0mm
                    \left.\ba{l}{\,}\\{\,}\ea\!\!\right]}}
\newcommand{\Hugebr}{{\arraycolsep 0mm
                    \left.\ba{l}{\,}\\{\,}\\{\,}\ea\!\!\right]}}
\newcommand{\HUgebr}{{\arraycolsep 0mm
                    \left.\ba{l}{\,}\\{\,}\\{\,}\vspace{-1mm}
                    \\{\,}\ea\!\!\right]}}

\newcommand{\hugecl}{{\arraycolsep 0mm
                    \left(\ba{l}{\,}\\{\,}\ea\right.\!\!}}
\newcommand{\Hugecl}{{\arraycolsep 0mm
                    \left(\ba{l}{\,}\\{\,}\\{\,}\ea\right.\!\!}}
\newcommand{\hugecr}{{\arraycolsep 0mm
                    \left.\ba{l}{\,}\\{\,}\ea\!\!\right)}}
\newcommand{\Hugecr}{{\arraycolsep 0mm
                    \left.\ba{l}{\,}\\{\,}\\{\,}\ea\!\!\right)}}

\newcommand{\hugepl}{{\arraycolsep 0mm
                    \left|\ba{l}{\,}\\{\,}\ea\right.\!\!}}
\newcommand{\Hugepl}{{\arraycolsep 0mm
                    \left|\ba{l}{\,}\\{\,}\\{\,}\ea\right.\!\!}}
\newcommand{\hugepr}{{\arraycolsep 0mm
                    \left.\ba{l}{\,}\\{\,}\ea\!\!\right|}}
\newcommand{\Hugepr}{{\arraycolsep 0mm
                    \left.\ba{l}{\,}\\{\,}\\{\,}\ea\!\!\right|}}

\newcommand{\MEq}[1]{\stackrel{
{\rm (#1)}}{=}}

\newcommand{\MLeq}[1]{\stackrel{
{\rm (#1)}}{\leq}}

\newcommand{\ML}[1]{\stackrel{
{\rm (#1)}}{<}}

\newcommand{\MGeq}[1]{\stackrel{
{\rm (#1)}}{\geq}}

\newcommand{\MG}[1]{\stackrel{
{\rm (#1)}}{>}}

\newcommand{\MPreq}[1]{\stackrel{
{\rm (#1)}}{\preceq}}

\newcommand{\MSueq}[1]{\stackrel{
{\rm (#1)}}{\succeq}}

\newcommand{\MSubeq}[1]{\stackrel{
{\rm (#1)}}{\subseteq}}

\newcommand{\MSupeq}[1]{\stackrel{
{\rm (#1)}}{\supseteq}}

\newcommand{\MRarrow}[1]{\stackrel{
{\rm (#1)}}{\Rightarrow}}

\newcommand{\MLarrow}[1]{\stackrel{
{\rm (#1)}}{\Leftarrow}}

\newcommand{\SZZpp}{
}

\newcommand{\vcc}{{c}^n}
\newcommand{\vck}{{k}^n}
\newcommand{\vcx}{{x}^n}
\newcommand{\vcy}{{y}^n}
\newcommand{\vcz}{{z}^n}
\newcommand{\vckone}{{k}_1^n}
\newcommand{\vcktwo}{{k}_2^n}
\newcommand{\vcxone}{{x}^n}

\newcommand{\vcxtwo}{{x}_2^n}
\newcommand{\vcyone}{{y}_1^n}
\newcommand{\vcytwo}{{y}_2^n}

\newcommand{\cvcx}{\check{x}^n}
\newcommand{\cvcy}{\check{y}^n}
\newcommand{\cvcz}{\check{z}^n}
\newcommand{\cvcxone}{\check{x}^n}
\newcommand{\cvcxtwo}{\check{x}_2^n}
\newcommand{\cvcxi}{\check{x}_i^n}

\newcommand{\hvcx}{\widehat{x}^n}
\newcommand{\hvcy}{\widehat{y}^n}
\newcommand{\hvcz}{\widehat{z}^n}
\newcommand{\hvckone}{\widehat{k}_1^n}
\newcommand{\hvcktwo}{\widehat{k}_2^n}

\newcommand{\hvcxone}{\widehat{x}^n}
\newcommand{\hvcxtwo}{\widehat{x}_2^n}
\newcommand{\hvcxi  }{\widehat{x}_i^n}

\newcommand{\lvcc}{{c}^n}
\newcommand{\lvck}{{k}^n}
\newcommand{\lvcx}{{x}^n}
\newcommand{\lvcy}{{y}^n}
\newcommand{\lvcz}{{z}^n}

\newcommand{\lvckone}{{k}_1^n}
\newcommand{\lvcktwo}{{k}_2^n}
\newcommand{\lvcxone}{{x}^n}

\newcommand{\lvcxtwo}{{x}_2^n}
\newcommand{\lvcyone}{{y}_1^n}
\newcommand{\lvcytwo}{{y}_2^n}

\newcommand{\clvcxone}{\check{x}^n}

\newcommand{\clvcxtwo}{\check{x}_2^n}

\newcommand{\hlvckone}{\widehat{k}_1^n}
\newcommand{\hlvcktwo}{\widehat{k}_2^n}

\newcommand{\hlvcxone}{\widehat{x}^n}

\newcommand{\hlvcxtwo}{\widehat{x}_2^n}

\newcommand{\rvcc}{{C}^n}
\newcommand{\rvck}{{K}^n}
\newcommand{\rvcx}{{X}^n}
\newcommand{\rvcy}{{Y}^n}
\newcommand{\rvcz}{{Z}^n}
\newcommand{\rvccone}{{C}_1^n}
\newcommand{\rvcctwo}{{C}_2^n}
\newcommand{\rvckone}{{K}_1^n}
\newcommand{\rvcktwo}{{K}_2^n}
\newcommand{\rvcxone}{{X}_1^n}

\newcommand{\rvcxtwo}{{X}_2^n}
\newcommand{\rvcyone}{{Y}_1^n}
\newcommand{\rvcytwo}{{Y}_2^n}
\newcommand{\hrvcx}{\widehat{X}^n}
\newcommand{\hrvcxone}{\widehat{X}_1^n}
\newcommand{\hrvcxtwo}{\widehat{X}_2^n}

\newcommand{\lrvcc}{{C}^n}
\newcommand{\lrvck}{{K}^n}
\newcommand{\lrvcx}{{X}^n}
\newcommand{\lrvcy}{{Y}^n}
\newcommand{\lrvcz}{{Z}^n}
\newcommand{\lrvckone}{{K}_1^n}
\newcommand{\lrvcktwo}{{K}_2^n}

\newcommand{\lrvcxone}{{X}^n}
\newcommand{\lrvcxtwo}{{X}_2^n}
\newcommand{\lrvcyone}{{Y}_1^n}
\newcommand{\lrvcytwo}{{Y}_2^n}
\newcommand{\rvcci}{{C}_i^n}
\newcommand{\rvcki}{{K}_i^n}
\newcommand{\rvcxi}{{X}_i^n}
\newcommand{\rvcyi}{{Y}_i^n}
\newcommand{\hrvcxi}{\widehat{X}_i^n}
\newcommand{\vcki}{{k}_i^n}
\newcommand{\vcsi}{{s}_i^n}
\newcommand{\vcti}{{t}_i^n}
\newcommand{\vcvi}{{v}_i^n}
\newcommand{\vcwi}{{w}_i^n}
\newcommand{\vcxi}{{x}_i^n}
\newcommand{\vcyi}{{y}_i^n}

\newcommand{\vcs}{{s}^n}
\newcommand{\vct}{{t}^n}
\newcommand{\vcv}{{v}^n}
\newcommand{\vcw}{{w}^n}
%
%

\newcommand{\SZZ}{

\newcommand{\vcc}{{\vc c}}
\newcommand{\vck}{{\vc k}}
\newcommand{\vcx}{{\vc x}}
\newcommand{\vcy}{{\vc y}}
\newcommand{\vcz}{{\vc z}}
\newcommand{\vckone}{{\vc k}_1}
\newcommand{\vcktwo}{{\vc k}_2}
\newcommand{\vcxone}{{\vc x}_1}
\newcommand{\vcxtwo}{{\vc x}_2}
\newcommand{\vcyone}{{\vc y}_1}
\newcommand{\vcytwo}{{\vc y}_2}

\newcommand{\cvcx}{\check{\vc x}}
\newcommand{\cvcy}{\check{\vc y}}
\newcommand{\cvcz}{\check{\vc z}}
\newcommand{\cvcxone}{\check{\vc x}_1}
\newcommand{\cvcxtwo}{\check{\vc x}_2}
\newcommand{\cvcxi}{\check{\vc x}_i}

\newcommand{\hvcx}{\widehat{\vc x}}
\newcommand{\hvcy}{\widehat{\vc y}}
\newcommand{\hvcz}{\widehat{\vc z}}
\newcommand{\hvckone}{\widehat{\vc k}_1}
\newcommand{\hvcktwo}{\widehat{\vc k}_2}
\newcommand{\hvcxone}{\widehat{\vc x}_1}
\newcommand{\hvcxtwo}{\widehat{\vc x}_2}

\newcommand{\lvcc}{{c}}
\newcommand{\lvck}{{k}}
\newcommand{\lvcx}{{x}}
\newcommand{\lvcy}{{y}}
\newcommand{\lvcz}{{z}}

\newcommand{\lvckone}{{k}_1}
\newcommand{\lvcktwo}{{k}_2}

\newcommand{\lvcxone}{{x}}

\newcommand{\lvcxtwo}{{x}_2}
\newcommand{\lvcyone}{{y}_1}
\newcommand{\lvcytwo}{{y}_2}

\newcommand{\clvcxone}{\check{x}_1}
\newcommand{\clvcxtwo}{\check{x}_2}

\newcommand{\hlvckone}{\widehat{k}_1}
\newcommand{\hlvcktwo}{\widehat{k}_2}
\newcommand{\hlvcxone}{\widehat{x}_1}
\newcommand{\hlvcxtwo}{\widehat{x}_2}

\newcommand{\rvcc}{{\vc C}}
\newcommand{\rvck}{{\vc K}}
\newcommand{\rvcx}{{\vc X}}
\newcommand{\rvcy}{{\vc Y}}
\newcommand{\rvcz}{{\vc Z}}
\newcommand{\rvccone}{{\vc C}_1}
\newcommand{\rvcctwo}{{\vc C}_2}
\newcommand{\rvckone}{{\vc K}_1}
\newcommand{\rvcktwo}{{\vc K}_2}

\newcommand{\rvcxone}{{\vc X}_1}

\newcommand{\rvcxtwo}{{\vc X}_2}
\newcommand{\rvcyone}{{\vc Y}_1}
\newcommand{\rvcytwo}{{\vc Y}_2}
\newcommand{\hrvcxone}{\widehat{\vc X}_1}
\newcommand{\hrvcxtwo}{\widehat{\vc X}_2}

\newcommand{\lrvcc}{{C}}
\newcommand{\lrvck}{{K}}
\newcommand{\lrvcx}{{X}}
\newcommand{\lrvcy}{{Y}}
\newcommand{\lrvcz}{{Z}}
\newcommand{\lrvckone}{{K}_1}
\newcommand{\lrvcktwo}{{K}_2}
\newcommand{\lrvcxone}{{X}_1}
\newcommand{\lrvcxtwo}{{X}_2}
\newcommand{\lrvcyone}{{Y}_1}
\newcommand{\lrvcytwo}{{Y}_2}
\newcommand{\rvcci}{{\vc C}_i}
\newcommand{\rvcki}{{\vc K}_i}
\newcommand{\rvcxi}{{\vc X}_i}
\newcommand{\rvcyi}{{\vc Y}_i}
\newcommand{\hrvcxi}{\widehat{\vc X}_i}
\newcommand{\vcki}{{\vc k}_i}
\newcommand{\vcsi}{{\vc s}_i}
\newcommand{\vcti}{{\vc t}_i}
\newcommand{\vcvi}{{\vc v}_i}
\newcommand{\vcwi}{{\vc w}_i}
\newcommand{\vcxi}{{\vc x}_i}
\newcommand{\vcyi}{{\vc y}_i}

\newcommand{\vcs}{{\vc s}}
\newcommand{\vct}{{\vc t}}
\newcommand{\vcv}{{\vc v}}
\newcommand{\vcw}{{\vc w}}
}
\newcommand{\One}{\rm (i)}
\newcommand{\Two}{\rm (ii)}
\newcommand{\Thr}{\rm (iii)}
\newcommand{\Fou}{\rm (iv)}
\newcommand{\Fiv}{\rm (v)}
\newcommand{\OnE}{1}

\newcommand{\pZKoKt}{p_{ZK_1K_2}}
\newcommand{\pZKoKtUn}{p^n_{ZK_1K_2}}

\section{Introduction \label{sec:introduction}}

In this paper, we consider the problem of strengthening the security of 
broadcasting secret sources encripted by common key criptsystems under 
the situation where the running criptsystems have some potential problems. 
More precisely, we consider two cryptosystems described as follows: two 
sources $X_1$ and $X_2$, respectively, are encrypted in a node to $C_1$ 
and $C_2$ using secret key $K_1$ and $K_2$. The cipher texts $C_1$ and 
$C_2$, respectively, are sent through public communication channels to 
the sink nodes 1 and 2. For each $i$, at the sink node $i$, $X_i$ is 
decrypted from $C_i$ using $K_i$. In this paper we assume we have two 
potentical problems in the above two cryptsystems. One is that the two 
common keys used in the above systems may have correlation. The other is 
that the adversary can use the side-channel, where the side information 
on two common keys can be obtained via the rate constraint noiseless 
channel. To solve this problem we formulate the post encryption coding 
system. In this communication system, we evaluate the information 
leakage on two secrete messages to the adversary. We provide an explicit 
sufficient condition for the information leakage to decay exponentially 
as the block length of encrypted source tends to infinity.

\section{Problem Formulation}

\subsection{Preliminaries}

In this subsection, we show the basic notations and related consensus 
used in this paper. 

\noindent{}
\textit{Random Source of Information and Key: \ }
For each $i=1,2$, let $X_i$ be a random variable from a finite set 
$\mathcal{X}_i$. For each $i=1,2$, let $\{X_{i,t}\}_{t=1}^\infty$ 
be two stationary discrete memoryless sources(DMS) 
such that for each $t=1,2,\ldots$, 
$X_{i,t}$ take values in finite set $\mathcal{X}_i$ 
and has  the same distribution as that of $X_i$ denoted by 
${p}_{X_i}=\{{p}_{X_i} (x_i)\}_{x_i \in \mathcal{X}_i}$.
The stationary DMS $\{X_{i,t}\}_{t=1}^\infty,$  
are specified with $p_{X_i}$. 

We next define the two keys used in the two common cryptosystems. 
For each $i=1,2$, let $(K_1,K_2)$ be a pair of two correlated 
random variables taken from the same finite set 
$\mathcal{X}_1\times \mathcal{X}_2$.
Let $\{(K_{1,t},K_{2,t}\}_{t=1}^\infty$ be a stationary discrete
memoryless source such that for each $t=1,2,\ldots$, $(K_{1,t},K_{2,t})$ 
takes values in $\mathcal{X}_1\times$ $\mathcal{X}_2$ 
and has the same distribution as that of $(K_1,K_2)$ denoted by 
$$
{p}_{K_1K_2}=\{{p}_{K_1K_2} (k_1,k_2)\}_{ 
  (k_1,k_2) \in \mathcal{X}_1 \times  \mathcal{X}_2}.
$$
The stationary DMS $\{(K_{1,t}, K_{2,t}\}_{t=1}^\infty$ 
is specified with ${p}_{K_1K_2}$. 
In this paper we assume that for each $i=1,2$, 
the marginal distribution 
$p_{K_i}$ is the uniform distribution over ${\cal X}_i$.  

\noindent{}\textit{Random Variables and Sequences: \ }
We write the sequence of random variables with length $n$ 
from the information sources as follows:
$X_i^n\defeq X_{i,1}X_{i,2}\cdots X_{i,n},i=1,2$. 
Similarly, the strings with length 
$n$ of $\mathcal{X}_i^n$ are written as 
${\vcxi}\defeq x_{i,1}x_{i,2}\cdots x_{i,n}\in\mathcal{X}_i^n$. 
For $({\vcxone}, {\vcxtwo})\in \mathcal{X}_1^n \times \mathcal{X}_2^n$, 
${p}_{X_1^nX_2^n}({\vcxone},{\vcxtwo})$ stands for the 
probability of the occurrence of 
$({\vcxone},{\vcxtwo})$. When the information source 
is memoryless specified with ${p}_{X_1X_2}$, we have 
the following equation holds:
$$
{p}_{X_1^nX_2^n}(x_1^n,x_2^n)
=\prod_{t=1}^n {p}_{X_1X_2}(x_{1,t},x_{2,t}).
$$
In this case we write ${p}_{X_1^nX_2^n}(x_1^n,x_2^n)$
as ${p}_{X_1X_2}^n(x_1^n,x_2^n)$. Similar notations are 
used for other random variables and sequences.

\noindent{}\emph{Consensus and Notations: }
Without loss of generality, throughout this 
paper, we assume that $X_1$ and $X_2$ are finite fields. The notation 
$\oplus$ is used to denote the field addition operation, while the 
notation $\ominus$ is used to denote the field subtraction operation, 
i.e., $a \ominus b = a \oplus (-b)$ for any elements $a, b$ from the 
same finite field. All discussions and theorems in this paper still hold 
although$X_1$ and $X_2$ are different finite fields. However, for the 
sake of simplicity, we use the same notation for field addition and 
subtraction for both $X_1$ and $X_2$. Throughout this paper all 
logarithms are taken to the base natural.

\subsection{Basic System Description}

In this subsection we explain the basic system setting and 
basic adversarial model we consider in this paper.
First, let the information source and the key be generated 
independently by three different parties
$\SgenOne$, $\SgenTwo$ and $\Kgen$ respectively.
In our setting, we assume the followings.
\begin{itemize}
	\item 	The random keys ${\rvckone}$  and ${\rvcktwo}$ 
                are generated by $\Kgen$
		from uniform distribution.
	\item	The key ${\rvckone}$ is correlated to  
		${\rvcktwo}$. 
	\item  	The sources ${\rvcxone}$ and ${\rvcxtwo}$ 
                are generated by $\Sgen$ and are correlated
		to each other.
	\item   The sources are independent to the keys.
\end{itemize}

Next, let the two correlated random sources ${\rvcxone}$ and 
${\rvcxtwo}$, respectively from $\SgenOne$ and $\SgenTwo$ be sent to 
two separated nodes $\mathsf{L}_1$ and $\mathsf{L}_2$. And let two random 
key (sources) ${\rvckone}$ and ${\rvcktwo}$ from $\Kgen$ be also sent 
separately to $\mathsf{L}_1$ and $\mathsf{L}_2$. Further settings of our 
system are described as follows. Those are also shown in Fig. \ref{fig:main}.
\begin{enumerate}
	\item \emph{Separate Sources Processing:} \ For each $i=1,2$, 
        at the node $\L_i$, $X_i^n$ is encrypted with 
        the key $K_i^n$ using the encryption function $\mathsf{Enc}_i$.
        The ciphertext ${\rvcci}$ of ${\rvcxi}$ is given by
        $$
        \rvcci \defeq \mathsf{Enc}_i(\rvcxi)=\rvcxi\oplus \rvcki.
        $$
	\item \emph{Transmission:} \ Next, the 
	ciphertexts ${\rvccone}$ and ${\rvcctwo}$, respectively 
        are sent to the information processing center $\D_1$ and $\D_2$ 
        through two \emph{public} communication channels. 
        Meanwhile, the keys ${\rvckone}$ and ${\rvcktwo}$, respectively 
        are sent to $\D_1$ and $\D_2$ through two \emph{private} 
        communication channels.
	\item \emph{Sink Nodes Processing:} \ For each $i=1,2$, in $\D_i$, 
        we decrypt the ciphertext ${\rvcci}$ using the key ${\rvcki}$ through 
        the corresponding decryption procedure $\mathsf{Dec}_i$ defined by 
	$\mathsf{Dec}_i({\rvcci})={\rvcci} \ominus {\rvcki}$. 
        It is obvious that we can correctly reproduce the source 
        output $\rvcx$ from $\rvcci$ and $\rvcki$ 
        by the decryption function $\mathsf{Dec}_i$.
\end{enumerate}

\newcommand{\Omit}{}
{
\begin{figure}[t]
\centering
\includegraphics[width=0.48\textwidth]{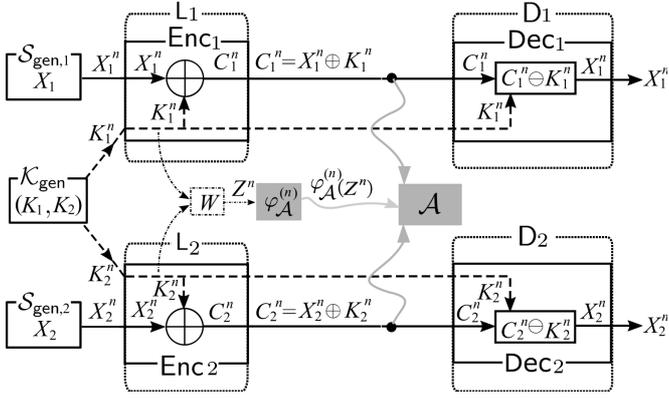}
\caption{Side-channel attacks to the two Shannon cipher systems.
\label{fig:main}}
\end{figure}
}
\noindent
\underline{\it Side-Channel Attacks by Eavesdropper Adversary:} 
An 
{\it adversary} $\A$ eavesdrops the public 
communication channel in the system. The adversary $\A$ 
also uses a side information obtained by side-channel attacks. 
Let ${\cal Z}$ be a finite set and let 
$W:$ ${\cal X}_1 \times {\cal X}_2\to {\cal Z}$ 
be a noisy channel. 
Let $Z$ be a channel output from $W$ for the input random variable $K$. 
We consider the discrete memoryless channel specified 
with $W$. Let $\rvcz \in {\cal Z}^n$ be a random variable obtained as 
the channel output by connecting $(\rvckone, \rvcktwo) 
\in {\cal X}_1^n \times {\cal X}_2^n$ to 
the input of channel. We write a conditional 
distribution on $\rvcz$ given $(\rvckone,\rvcktwo)$ as 
$$
W^n=\left\{W^n(\vcz|\vckone,\vcktwo)\right
\}_{(\vckone,\vcktwo,\lvcz) \in 
       {\cal X}_1^n 
\times {\cal X}_2^n 
\times {\cal Z}^n}.
$$
Since the channel is memoryless, we have 
\beq
W^n({\vcz}|\vckone,\vcktwo)=\prod_{t=1}^nW (z_t|k_{1,t},k_{2,t}).
\label{eqn:sde0}
\eeq
On the above output $\rvcz$ of $W^n$ for the input 
$(\rvckone,\rvcktwo)$, we assume the followings.
\begin{itemize}
\item The two random pairs $(X_1,X_2)$, $(K_1,K_2)$ 
and the random variable $Z$, satisfy 
$(X_1,X_2) \perp (K_1,K_2, Z)$, which implies that 
$(X_1^n,X_2^n) \perp (K_1^n,K_2^n, Z^n)$.
\item $W$ is given in the system and the adversary ${\cal A}$ 
can not control $W$. 
\item By side-channel attacks, the adversary ${\cal A}$ 
can access $Z^n$. 
\end{itemize}
We next formulate side information the adversary ${\cal A}$ 
obtains by side-channel attacks. For each $n=1,2,\cdots$, 
let $\varphi_{\cal A}^{(n)}:{\cal Z}^n 
\to {\cal M}_{\cal A}^{(n)}$ be an encoder function. 
Set 
$
\varphi_{\cal A} \defeq \{\varphi_{\cal A}^{(n)}\}_{n=1,2,\cdots}.
$ 
Let 
$$ 
R_{\cal A}^{(n)}\defeq 
\frac{1}{n} \log ||\varphi_{\cal A}||
=\frac{1}{n} \log |{\cal M}_{\cal A}^{(n)}|
$$
be a rate of the encoder function $\varphi_{\cal A}^{(n)}$. 
For $R_{\cal A}>0$, we set 
$$
{\cal F}_{\cal A}^{(n)}(R_{\cal A})
\defeq \{ \varphi_{\cal A}^{(n)}: R_{\cal A}^{(n)} \leq R_{\cal A}\}.
$$ 
On encoded side information the adversary ${\cal A}$ obtains we assume 
the following.
\begin{itemize}
\item The adversary ${\cal A}$, having accessed $Z^n$, obtains 
the encoded additional information $\varphi_{\cal A}^{(n)}(\rvcz)$.
For each $n=1,2,\cdots$, the adversary ${\cal A}$ 
can design $\varphi_{\cal A}^{(n)}$. 
\item The sequence $\{R_{\cal A}^{(n)}\}_{n=1}^{\infty}$
must be upper bounded by a prescribed value. 
In other words, the adversary ${\cal A}$ must use $ \varphi_{\cal A}^{(n)}$ 
such that for some $R_{\cal A}$ and for any sufficiently large $n$, 
$\varphi_{\cal A}^{(n)}\in {\cal F}_{\cal A}^{(n)}(R_{\cal A})$. 
\end{itemize}


\begin{figure}[t] 
	\centering 
	\includegraphics[width=0.48\textwidth]
	{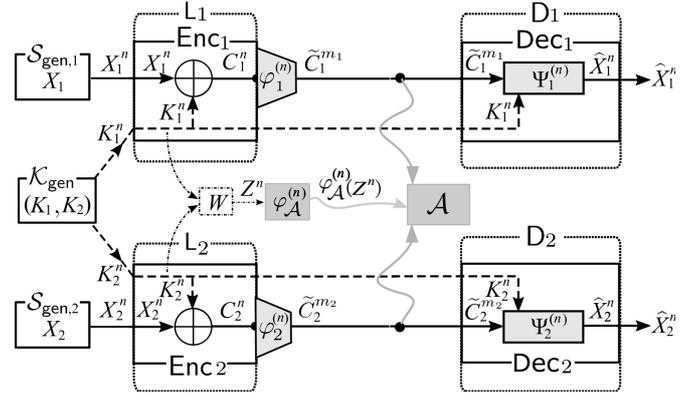}
	\caption{Post-encryption coding system.
\label{fig:Gensolution}}%
\end{figure}

As a soultion to the side channel attacks, we consider 
the post-encryption coding system. This system is shown 
in Fig. \ref{fig:Gensolution}.  

\begin{enumerate}
	\item\emph{Encoding at Source node $\L_i,i=1,2$:} \  For each $i=1,2$, 
        we first use 
	$\varphi_i^{(n)}$ to encode the ciphertext 
        $C_i^{n}=X_i^{n}\oplus K_i^{n}$.
        Formal definition of $\varphi_i^{(n)}$ is 
        $\varphi_i^{(n)}:$ ${\cal X}_i^n \to $${\cal X}_i^{m_i}$. 
	Let $\widetilde{C}_i^{m_i}=\varphi_i^{(n)}({\rvcci})$.
	Instead of sending ${\rvcci}$, we send $\tilde{C}_i^{m_i}$ 
        to the public communication channel. 
        \item\emph{Decoding at Sink Nodes $\D_i,i=1,2$: \ }
        For each $i=1,2$, $\D_i$ receives $\widetilde{C}_i^{m_i}$ 
        from public communication channel. 
        Using common key $\rvcki$ and the decoder function 
        $\Psi_{i}^{(n)}: {\cal X}_i^m \times {\cal X}_i^n 
        \to {\cal X}_i^n$, $\D_i$ outputs an estimation 
        $\hat{X}_i^n= \Psi_i^{(n)}(\tilde{C}_i^{m_i},\rvcki)$ 
        of $\rvcxi$.   
\end{enumerate}

\noindent
\underline{\it On Reliability and Security:}
From the description of our system in the previous section,
the decoding process in our system above is successful if 
$\widehat{X}^{n}=X^{n}$ holds.
Combining this and (\ref{eq:source_estimation}), 
it is clear that the decoding error probabilities 
$p_{{\rm e},i},i=1,2,$ are as follows:
\begin{align}
p_{{\rm e},i}=& 
p_{{\rm e}}(\varphi_i^{(n)},\Psi_i^{(n)}|{p}_{X_i}^n)
     \defeq \Pr[\Psi_i^{(n)}(\varphi_i^{(n)}({\rvcxi}))
\neq {\rvcxi}]. 
\notag
\end{align}
Set $M_{\cal A}^{(n)}=\varphi_{\cal A}^{(n)}(Z^n)$. 
The information leakage $\Delta^{(n)}$ on $(X_1^n,X_2^n)$ from 
$(\widetilde{C}_1^{m_1},\widetilde{C}_2^{m_2},M_{\cal A}^{(n)})$
is measured by the mutual information 
between $(X_1^n,X_2^n)$ and 
$(\widetilde{C}_1^{m_1},
  \widetilde{C}_2^{m_2},$ 
$M_{\cal A}^{(n)})$. This quantity 
is formally defined by 
\begin{align*}
&\Delta^{(n)}=
\Delta^{(n)}(\varphi_1^{(n)},\varphi_2^{(n)},\varphi_{\cal A}^{(n)}|
{p}_{X_1X_2}^n, {\pZKoKtUn})
\\
&\defeq I(X_1^nX_2^n;\widetilde{C}_1^{m_2}, 
                     \widetilde{C}_2^{m_2},
                     M_{\cal A}^{(n)}).  
\end{align*}

\noindent
\underline{\it Reliable and Secure Framework:} \ 

\begin{definition}
        A pair $(R_1,R_2)$ is achievable under $R_{\cal A}$ $>0$
        for the system $\mathsf{Sys}$ if there exists 
        two sequences 
        $\{(\varphi_i^{(n)},$ $\Psi_i^{(n)})\}_{n \geq 1},i=1,2,$
        such that 
        $\forall \epsilon>0$, $\exists n_0=n_0(\epsilon) \in\mathbb{N}_0$, 
	$\forall n\geq n_0$, we have for $i=1,2,$
	\begin{align*}
        &\frac{1}{n} \log |{\cal X}_i^{m_i}| 
        = \frac{m_i}{n} \log |{\cal X}_i| \leq R_i,\\  				
        &p_{{\rm e}}(\varphi_i^{(n)},\Psi_i^{(n)}|p^n_{X_i})\leq \epsilon,
	\end{align*}
	and for any eavesdropper $\A$ with $\varphi_{\A}$ satisfying 
        $\varphi_{\cal A}^{(n)}\in {\cal F}_{\cal A}^{(n)}(R_{\cal A})$, 
        we have 
	\begin{align*}
        \Delta^{(n)}(\varphi_1^{(n)}, 
                     \varphi_2^{(n)},
                     \varphi_{\cal A}^{(n)}
        |{p}_{X_1X_2}^n,{\pZKoKtUn})
	\leq \epsilon.
	\end{align*}
\end{definition}

\begin{definition}{\bf (Reliable and Secure Rate Region)}
        Let $\mathcal{R}_{\mathsf{Sys}}({p}_{X_1X_2},$ ${p}_{K_1K_2},W)$
	denote the set of all $(R_{\cal A},R)$ such that
        $R$ is achievable under $R_{\cal A}$. 
        We call $\mathcal{R}_{\mathsf{Sys}}({p}_{X_1X_2},{p}_{K_1K_2},$ $W)$
        the \emph{\bf reliable and secure rate} region. 
\end{definition}

\begin{definition}
        A five tuple $(R_1,R_2,E_1,E_2, F)$ is achievable under \\ 
        $R_{\cal A}>0$
        for the system $\mathsf{Sys}$ if there exists 
        a sequence $\{(\varphi_i^{(n)},$ $\Psi_i^{(n)})\}_{n \geq 1}$, 
        $i=1,2$, such that $\forall \epsilon>0$, 
        $\exists n_0=n_0(\epsilon)\in\mathbb{N}_0$, 
	$\forall n$ $\geq n_0$, we have for $i=1,2,$
	\begin{align*}
        &\frac {1}{n} \log |{\cal X}_i^{m_i}|=\frac {m_i}{n}
         \log |{\cal X}_i| \leq R_i,
        \\
        & p_{{\rm e}}(\varphi_i^{(n)},\psi_i^{(n)}|{p}_{X_i}^n) 
        \leq {\ExP}^{-n(E_i-\epsilon)},
	\end{align*}
	and for any eavesdropper $\A$ with $\varphi_{\A}$
        satisfying $\varphi_{\cal A}^{(n)} 
        \in {\cal F}_{\cal A}^{(n)}(R_{\cal A})$, 
        we have 
	\begin{align*}
	\Delta^{(n)}(\varphi_1^{(n)},
                     \varphi_2^{(n)},
                     \varphi_{\cal A}^{(n)}
        |{p}_{X_1X_2}^n,{\pZKoKtUn})
        \leq {\ExP}^{-n(F-\epsilon)}.
	\end{align*}
\end{definition}

\begin{definition}{\bf (Rate, Reliability, and Security Region)}
        Let $\mathcal{D}_{\mathsf{Sys}}({p}_{X_1X_2},$ ${p}_{K_1K_2},W)$
	denote the set of all $(R_{\cal A},R,E,F)$ such that
        $(R_1,R_2,E_1,E_2,F)$ is achievable under $R_{\cal A}$. 
        We call $\mathcal{D}_{\mathsf{Sys}}({p}_{X_1X_2},$ ${p}_{K_1K_2},W)$
        the \emph{\bf rate, reliability, and security} region.
\end{definition}

\begin{figure}[t] 
	\centering 
	\includegraphics[width=0.48\textwidth]
	{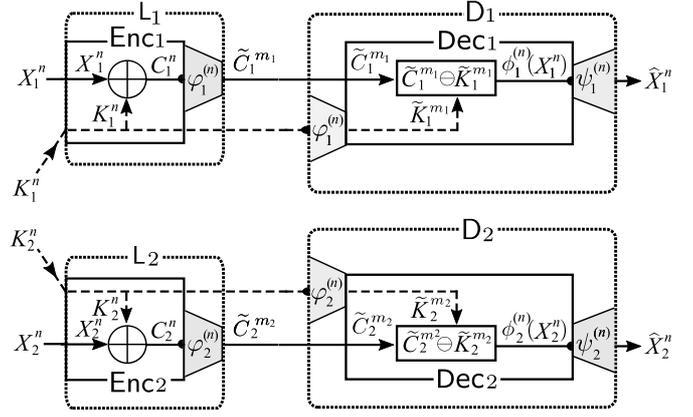}
	\caption{Our proposed solution: linear encoders 
	as privacy amplifiers.
\label{fig:solution}}%
\end{figure}
\section{Proposed Idea: Affine Encoder as Privacy Amplifier}
%
For each $n=1,2,\cdots$, let 
$\phi_i^{(n)}: {\cal X}_i^n \to {\cal X}_i^{m_i}$ 
be a linear mapping. We define the mapping $\phi_i^{(n)}$ by 
\beq
\phi_i^{(n)}({\vcxi})
={\vcxi} A_i \mbox{ for }{\vcxi} \in {\cal X}_i^n,
\label{eq:homomorphica}
\eeq
where $A_i$ is a matrix with $n$ rows and $m_i$ columns. 
Entries of $A_i$ are from ${\cal X}_i$. 
We fix $b_i^{m_i}\in \mathcal{X}_i^{m_i}$. 
Define the mapping 
$\varphi_i^{(n)}: {\cal X}_i^n \to {\cal X}_i^{m_i}$
by 
\begin{align}
\varphi_i^{(n)}({\vcki}):=&
\phi_i^{(n)}({\vcki})\oplus b_i^{m_i}
\notag \\
=&{\vcki} A_i\oplus b_i^{m_i}, 
\mbox{ for }{\vcki} \in \mathcal{X}_i^n.
\label{eq:homomorphic}
\end{align}
The mapping $\varphi_i^{(n)}$ is called the affine 
mapping induced by the linear mapping 
$\phi_i^{(n)}$ and constant vector 
$b_i^{m_i}$ $\in{\cal X}_i^{m_i}$.
By the definition (\ref{eq:homomorphic}) of 
$\varphi_i^{(n)}$, those satisfy the following 
affine structure: 
\begin{align}
&\varphi_i^{(n)}({\vcxi} \oplus {\vcki})
({\vcxi}\oplus {\vcki})A_i\oplus b_i^{m_i}=
{\vcxi}A_i\oplus({\vcki}A_i\oplus b_i^{m_i})
\notag\\
&=\phi_i^{(n)}({\vcxi})\oplus \varphi_i^{(n)}({\vcki}),
\mbox{ for } {\vcxi}, {\vcki} \in {\cal X}_i^n.
\label{eq:affine}
\end{align}
Next, let $\psi_i^{(n)}$ be the corresponding decoder for 
$\phi_i^{(n)}$ such that
$
\psi_i^{(n)}: \mathcal{X}_i^{m_i} 
\rightarrow \mathcal{X}_i^{n}.
$
Note that $\psi_i^{(n)}$ does not have a linear structure in general. 

\noindent
\underline{\it Description of Proposed Procedure:} 
We describe the procedure of our privacy 
amplified system as follows. 
\begin{enumerate}
	\item\emph{Encoding at Source node $\L_i,i=1,2$:} \ First, 
        we use $\varphi_i^{(n)}$ to encode the ciphertext 
        $C_i^{n}=X_i^{n}\oplus K_i^{n}$
	Let $\widetilde{C}_i^{m_i}=\varphi_i^{(n)}({\rvcci})$.
	Then, instead of sending ${\rvcc}$, we send $\tilde{C}_i^{m_i}$ 
        to the public communication channel. By the affine 
        structure (\ref{eq:affine}) of encoder we have that
        \begin{align}
        &\widetilde{C}_i^{m_i}=\varphi_i^{(n)}({\rvcxi}\oplus {\rvcki})
        \notag \\
        &=\phi_i^{(n)}({\rvcxi}) \oplus \varphi_i^{(n)}({\rvcki})
        =\widetilde{X}_i^{m_i} \oplus \widetilde{K}_i^{m_i},
        \label{eqn:aSdzx} 
        \end{align}
        where we set 
        $
	\widetilde{X}_i^{m_i} \defeq \phi_i^{(n)}({\rvcxi}),
        \widetilde{K}_i^{m_i} \defeq \varphi_i^{(n)}({\rvcki}).
        $
	\item\emph{Decoding at Sink Node $\D_i,i=1,2$: \ }
	First, using the linear encoder
	$\varphi_i^{(n)}$,
	$\D_i$ encodes the key $\rvcki$ received
	through private channel into
	$\widetilde{K}_i^{m_i}=$$\varphi_i^{(n)}({\rvcki})$.
	Receiving $\widetilde{C}_i^{m_i}$ from
	public communication channel, $\D_i$ computes
	$\widetilde{X}_i^{m_i}$ in the following way.
        From (\ref{eqn:aSdzx}), we have that 
        the decoder $\D_i$ can obtain 
        $\widetilde{X}_i^{m_i}$ $=\phi_i^{(n)}({\rvcxi})$
        by subtracting 
        $\widetilde{K}_i^{m_i}=\varphi_i^{(n)}({\rvcki})$ 
        from $\widetilde{C}_i^{m_i}$. 
	Finally, $\D_i$ outputs $\hrvcxi$
	by applying the decoder 
        $\psi_i^{(n)}$ to $\widetilde{X}_i^{m_i}$ as follows:
	\begin{align}
		\hrvcxi &=\psi_i^{(n)}(\widetilde{X}_i^{m_i}) 
         	         =\psi_i^{(n)}(\phi_i^{(n)}({\rvcxi})). 
                  \label{eq:source_estimation}
	\end{align}
\end{enumerate}
Our privacy amplified system described above is illustrated 
in Fig. \ref{fig:solution}.
\
\section{Main Results}

In this section we state our main results. To describe our results we define
several functions and sets. Let $U$ be an auxiliary random variable 
taking values in a finite set ${\cal U}$. We assume that the joint 
distribution of $(U,Z,K_1,K_2)$ is 
$$ 
p_{UZK_1K_2}(u,z,k_1,k_2)=p_{U}(u)
p_{{Z}|U}(z|u)p_{K_1K_2|Z}(k_1,k_2|z).
$$
The above condition is equivalent to 
$U \markov Z \leftrightarrow (K_1,K_2)$. 
In the following argument for convenience 
of descriptions of definitions we use 
the following notations: 
\begin{align*}
& R_3 \defeq R_1+R_2, 
{\cal X}_3 \defeq {\cal X}_1\times {\cal X}_2,
\\ 
& k_3 \defeq (k_1,k_2), K_3 \defeq (K_1,K_2). 
\end{align*}
For each $i=1,2,3$, we simply write $p_i=p_{UZK_i}$. 
Specifically, for $i=3$, we have $p_3=p_{UZK_1K_2}=p$.  
Define the three sets of probability distribution 
$p=p_{UZK_1K_2}$ by 
\begin{align*}
{\cal P}(p_{ZK_i})\defeq& 
\{p_{UZK_i}: \pa {\cal U} \pa \leq \pa {\cal Z} \pa+1,
U \markov Z \markov K_i \},
\\
&i=1,2,3.
\end{align*}
For $i=1,2,3$, set
\begin{align*}
&{\cal R}_i(p_i)
\defeq 
\ba[t]{l}
\{(R_{\cal A},R_i):R_{\cal A}, R_i \geq 0,
\vSpa\\
\: R_{\cal A}\geq I_{\empty}({Z};{U}), 
   R_i \geq H_{\empty}({K}_i|{U})\},
\ea
\\
&{\cal R}_i(p_{ZK_i}) \defeq \bigcup_{p_i \in {\cal P}(p_{ZK_i})}
{\cal R}_i(p_i).
\end{align*}

The two regions ${\cal R}_i(p_{ZK_i}), i=1,2$ have the same form  
as the region appearing as the admissible rate region 
in the one-helper source coding problem posed and investigated 
by Ahlswede and K\"orner \cite{ahlswede:75}. 

We can show that the region ${\cal R}_i(p_{ZK_i}), i=1,2,$ 
and ${\cal R}_3({\pZKoKt})$ 
satisfy the following property.
\begin{pr}\label{pr:pro0xx}  
$\quad$
\begin{itemize}
\item[a)] 
The region ${\cal R}(p_{ZK_i}),i=1,2$, is a closed convex 
subset of $\mathbb{R}_{+}^2$.
The region ${\cal R}_3({\pZKoKt})$ is a closed convex 
subset of $\mathbb{R}_{+}^3$.
\item[b)] The bound $|{\cal U}|\leq |{\cal Z}|+1$
is sufficient to describe ${\cal R}_i(p_{ZK_i}),i=1,2,3$. 
\end{itemize}
\end{pr}

\newcommand{\OmitZZ}{}{
We next explain that the region ${\cal R}_i(p_{ZK_i}),i=1,2,$ 
and ${\cal R}_3({\pZKoKt})$ can be expressed with 
a family of supporting hyperplanes. To describe this result 
we define three sets of probability distributions on 
${\cal U}$ $\times{\cal Z}$ $\times{\cal X}_1$$\times{\cal X}_2$ 
by
\begin{align*}
\tilde{\cal P}(p_{ZK_i}) \defeq &
 \{p=p_{UZK_i}: \pa {\cal U} \pa \leq \pa {\cal X} \pa,
  U \markov  Z\markov K_i \},
\\
  & i=1,2,3.
\end{align*}
For $i=1,2,3$, and $\mu\in [0,1]$, define
\begin{align*}
& R^{(\mu)}(p_{ZK_i})
\defeq 
\min_{p \in \tilde{\cal P}(p_{ZK_i})}
\left\{ \mu I_p(Z;U)+ \bar{\mu}H_p(K_i|U)\right\},
\\
\end{align*}
Furthermore, for $i=1,2,3$, define
\begin{align*}
& {\cal R}_{{\rm sh},i}(p_{ZK_i})
  \defeq \bigcap_{\mu \in [0,1]}\{(R_{\cal A},R_i):
\ba[t]{l}
 \mu R_{\cal A}+ \bar{\mu}R_i \\ 
\geq R^{(\mu)}(p_{ZK_i})\}.
\ea 
\end{align*}
Then we have the following property.
\begin{pr}\label{pr:pro0zc} $\quad$
\begin{itemize}
\item[a)] The bound $|{\cal U}|\leq |{\cal Z}|$
is sufficient to describe 
$R_i^{(\mu)}($ $p_{ZK_i}), 
   i=1,2$, and 
$R_3^{(\mu)}({\pZKoKt})$. 

\item[b)] For every $\mu\in [0,1]$, we have 
\begin{align}
&\min_{(R_{\cal A},R_i)\in {\cal R}(p_{ZK_i})}
\{\mu R_{\cal A}+\bar{\mu}R_i\}
\notag\\
&=R^{(\mu)}(p_{ZK_i}),i=1,2,
\label{eqn:SDAzz}
\\
&\min_{(R_{\cal A},R_1,R_2)\in {\cal R}({\pZKoKt})}
\{\mu R_{\cal A}+\bar{\mu}(R_1+R_2)\}
\notag\\
&=R^{(\mu)}({\pZKoKt}).
\label{eqn:SDAzzb}
\end{align}
\item[c)] For any ${\pZKoKt}$ we have
\begin{align}
&{\cal R}_{{\rm sh},i}(p_{ZK_i})
={\cal R}(p_{ZK_i}),i=1,2, 
\label{eqn:PropEqBa}\\
& {\cal R}_{{\rm sh},3}({\pZKoKt})
 ={\cal R}_3({\pZKoKt}).
\label{eqn:PropEqB}
\end{align}
\end{itemize}
\end{pr}
}


We define several quantities to state a result on 
$\mathcal{D}_{\mathsf{Sys}}($ ${p}_{X_1X_2},{p}_{K_1K_2},W)$. 
Let $i\in \{ 1,2 \}$. We first define a function related to an 
exponential upper bound of 
$p_{{\rm e}}(\phi_i^{(n)},\psi_i^{(n)}|{p}_{X_i}^n)$. 
Let $\overline{X}_i$ be an arbitrary random variable
over $\mathcal{X}_i$ and has a probability distribution 
$p_{\overline{X}_i}$. Let $\mathcal{P}(\mathcal{\cal X}_i)$ 
denote the set of all probability distributions on 
$\mathcal{X}_i$. For $ R_i \geq 0$ and $p_{{X}_i}\in $ 
$\mathcal{P}( \mathcal{\cal X}_i)$, we 
define the following function:
\begin{align*}
	E(R_i|p_{{X}_i}) &:{=}
	\min_{ p_{ \overline{X}_i} \in 
	\mathcal{P}(\mathcal{\cal X}_i)}
	\{[R_i- H(\overline{X}_i)]^{+}
	   +D(p_{ \overline{X}_i}||p_{X_i})\}.
\end{align*}
We next define a function related to an exponential 
upper bound of 
$\Delta^{(n)}(\varphi_1^{(n)},\varphi_2^{(n)},
\varphi_{\A}^{(n)}|{p}_{X_1X_2}^n,{\pZKoKtUn})$. 
For each $i=1,2,3$, we define a set of probability distributions on 
${\cal U}$ $\times{\cal Z}$ $\times {\cal X}_i$ by
\begin{align*}
{\cal Q}(p_{K_i|Z}) \defeq 
       & \{q_i=q_{UZK_i}: q_{K_iZ|U}=p_{K_iZ|U} \\
       &\:\mbox{ for some }p_i\in \tilde{\cal P}(p_{ZK_i})\}. 
\end{align*}
For each $i=1,2,3$, for $(\mu,\alpha) \in [0,1]^2$, and 
for $q_i=q_{UZK_i}\in {\cal Q}(p_{K_i|Z})$, 
define 
\begin{align*}
& \omega_{q_i|p_Z}^{(\mu,\alpha)}(z,k_i|u)
  \defeq \bar{\alpha} \log \frac{q_Z(z)}{p_Z(z)}
\\
&+ \alpha \left[
  {\mu}\log \frac{q_{Z|U}(z|u)}{p_{Z}(z)}\right.
\left. +\bar{\mu}\log \frac{1}{q_{K_i|U}(k_i|u)}
\right],
\\
& \Omega^{(\mu,\alpha)}(q_i|p_{Z})
\defeq 
-\log {\rm E}_{q}
\left[\exp\left\{-\omega^{(\mu,\alpha)}_{q_i|p_Z}(Z,K_i|U)
     \right\}\right],
\\
&\Omega^{(\mu,\alpha)}(p_{ZK_i})
\defeq 
\min_{\scs 
   \atop{\scs 
    q_i \in {{\cal Q}}(p_{K_i|Z})
   }
}
\Omega^{(\mu,\alpha)}(q_i|p_Z),
\\
& F^{(\mu,\alpha)}({\mu}R_{\cal A}+\bar{\mu} R_{i}|p_{ZK_i})
\\
&\defeq \frac{\Omega_i^{(\mu,\alpha)}(p_{K_i},W)
-\alpha({\mu}R_{\cal A} + \bar{\mu} R_{i})}
{2+\alpha \bar{\mu}},
\\
& F(R_{\cal A},R_{i}|p_{ZK_i})
\\
&\defeq \sup_{ (\mu,\alpha)\in [0,1]^2}
F^{(\mu,\alpha)}({\mu}R_{\cal A}+\bar{\mu} R_{i}|p_{ZK_i}).
\end{align*}
We next define a function serving as a lower 
bound of $F(R_{\cal A},R_{i}|p_{ZK_i}),i=1,2,3$. 
For each $i=1,2,3,$ and for each 
$p_{i}\in \tilde{\cal P}(p_{ZK_i})$, 
define 
\begin{align*}
& \tilde{\omega}_{p_i}^{(\mu)}({z},k_i|u)
\defeq {\mu}\log \frac{p_{{Z}|U}({z}|u)}{p_Z(z)}
  +\bar{\mu}\log \frac{1}{p_{K_i|U}(K_i|U)},
\\
& \tilde{\Omega}^{(\mu,\lambda)}(p_i)
\defeq 
-\log 
{\rm E}_{p}
\left[\exp\left\{-\lambda
\tilde{\omega}_{p_i}^{(\mu)}({Z},K_i|U)\right\}\right].
\end{align*}
Furthermore, set
\begin{align*}
& \tilde{\Omega}^{(\mu,\lambda)}(p_{ZK_i})
 \defeq \min_{\scs \atop{\scs 
  p_i\in {\tilde{\cal P}(p_{ZK_i})}}}
\tilde{\Omega}^{(\mu,\lambda)}(p_i),
\\
& \tilde{F}^{(\mu,\lambda)}
  ({\mu}R_{\cal A}+\bar{\mu}R_{i}|p_{ZK_i}) 
\\
&\defeq 
\frac{\tilde{\Omega}^{(\mu,\lambda)}(p_{ZK_i})
-\lambda({\mu}R_{\cal A}+\bar{\mu}R_{i})}{2+\lambda(5-{\mu})},
\\
& \tilde{F}(R_{\cal A},R_i|p_{ZK_i})
\defeq \sup_{\scs \lambda \geq 0, \atop{\scs \mu \in [0,1]}} 
\tilde{F}^{(\mu,\lambda)}
({\mu} R_{\cal A}+ \bar{\mu}R_i|p_{ZK_i}).
\end{align*}

We can show that the above functions satisfy the 
following property. 
\begin{property}\label{pr:pro1}  
$\quad$
\begin{itemize}
\item[a)]
For each $i=1,2,3$, the cardinality bound 
$|{\cal U}|\leq |{\cal {Z}}|$ in ${\cal Q}(p_{K_i|{Z}})$
is sufficient to describe the quantity
$\Omega^{(\mu,\alpha)}(p_{ZK_i})$. 
Furthermore, the cardinality bound 
$|{\cal U}|\leq |{\cal {Z}}|$ in 
${\cal P}_{\rm sh}(p_{ZK_1K_2})$
is sufficient to describe the quantity
$\tilde{\Omega}^{(\mu,\lambda)}(p_{ZK_1K_2})$. 

\item[b)] For $i=1,2,3$ and for any $R_{\cal A},R_i \geq0$, we have 
\beqno
& &F(R_{\cal A},R_{\empty}|p_{ZK_i})\geq 
\tilde{F}(R_{\cal A},R_{i}|p_{ZK_i}).
\eeqno

\item[c)] For $i=1,2,3$ and for any 
$p_i \in {\cal P}_{\rm sh}(p_{ZK_i})$
and any $(\mu,\lambda)\in [0,$ $1]^2$, we have 
\beq
0\leq \tilde{\Omega}^{(\mu,\lambda)}(p_i) 
 \leq \mu \log |{\cal Z}|
    + \bar{\mu} \log |{\cal K}_i|.
\label{eqn:Asddx}
\eeq

\item[d)] Fix any $p=p_{U{Z}K}\in {\cal P}_{\rm sh}(p_{K},W)$ 
and $\mu\in [0,1]$. 
For $\lambda \in [0,1]$, we define a probability distribution 
$p_i^{(\lambda)} = p_{UZK_i}^{(\lambda)}$ by
\begin{align*} 
& p_i^{(\lambda)}(u,{z},k_i)
\defeq
\frac{
p_i(u,{z},k_i)
\exp\left\{
-\lambda \tilde{\omega}^{(\mu)}_{p_i}(z,k_i|u)
\right\}
}
{{\rm E}_{p_i}
\left[\exp\left\{-\lambda
\tilde{\omega}^{(\mu)}_{p_i}({Z},K_i|U)\right\}\right]}.
\end{align*}
Then for each $i=1,2,3$ and for $\lambda \in [0,1/2]$, 
$\tilde{\Omega}^{(\mu,\lambda)}(p_i)$ is 
twice differentiable. Furthermore, 
for $\lambda \in [0,1/2]$, we have
\beqno
& & \frac{\rm d}{{\rm d}\lambda}
\tilde{\Omega}^{(\mu,\lambda)}(p_i)
={\rm E}_{{p_i}^{(\lambda)}}
\left[\tilde{\omega}_{p_i}^{(\mu)}({Z},K_i|U)\right],
\\
& & \frac{\rm d^2}{{\rm d}\lambda^2} 
\tilde{\Omega}^{(\mu,\lambda)}(p_i)
=-{\rm Var}_{p_i^{(\lambda)}}
\left[\tilde{\omega}^{(\mu)}_{p_i}({Z},K_i|U)\right].
\eeqno
The second equality implies that 
$\tilde{\Omega}^{(\mu,\lambda)}(p_i|p_{ZK_i})$ 
is a concave function of $\lambda\geq0$. 
\item[e)] 
For $(\mu,\lambda)\in [0,1] \times [0,1/2]$, define 
\begin{align*}
& \rho^{(\mu,\lambda)}(p_{ZK_i})
\\
&\defeq {\max_{\scs (\nu, p_i) \in [0,\lambda] 
    \atop{\scs \times \tilde{\cal P}(p_{ZK_i}):
        \atop{\scs \tilde{\Omega}^{(\mu,\lambda)}(p_i) 
             \atop{\scs 
             =\tilde{\Omega}^{(\mu,\lambda)}(p_{ZK_i})
             }}}}
}
{\rm Var}_{{p_i^{(\nu)}}}
\left[\tilde{\omega}^{(\mu)}_{p_i}({Z},K_i|U)\right],
\end{align*}
and set
\begin{align*}
& 
\rho(p_{ZK_i})
\defeq \max_{(\mu,\lambda)\in [0,1] \times [0,1/2]}
\rho^{(\mu,\lambda)}(p_{ZK_i}).
\end{align*}
Then we have $\rho(p_{ZK_i})<\infty$. Furthermore, 
for any $(\mu,\lambda) \in [0,1]\times[0,1/2]$, we have
$$
\tilde{\Omega}^{(\mu,\lambda)}(p_{ZK_i}) 
\geq \lambda R^{(\mu)}(p_{ZK_i}) 
-\frac{\lambda^2}{2}\rho(p_{ZK_i}).
$$
\item[f)] For every $\tau \in(0,(1/2)\rho(p_{ZK_i})$,
the condition 
$(R_{\cal A},$ $R_{\empty}+\tau) \notin {\cal R}(p_{ZK_i})$
implies 
\begin{align*}
& \tilde{F}(R_{\cal A},R_{\empty}| p_{ZK_i})
> \ts \frac{\rho(p_{ZK_i})}{4} \cdot g^2
\left({\ts \frac{\tau}{\rho(p_{ZK_i})}}\right)>0,
\end{align*}
where $g$ is the inverse function of 
$\vartheta(a) \defeq a+(5/4)a^2, a \geq 0$.
\end{itemize}
\end{property}

Proof of this property is found in 
Oohama \cite{oohama2015exponent}(extended version).
We set 
\begin{align*}
&F_{\min}(R_{\A},R_1,R_2|{\pZKoKt}) 
\defeq \min_{i=1,2,3} F(R_{\A},R_i|p_{ZK_i}). 
\end{align*}
Our main result is as follows. 
\begin{theorem}\label{Th:mainth2}{
\rm For any $R_{\cal A}, R_1,R_2>0$ and 
any ${\pZKoKt}$, there exists two sequence of mappings 
$\{(\varphi_i^{(n)}, \psi_i^{(n)}) \}_{n=1}^{\infty},i=1,2$
such that for any $p_{X_i},i=1,2,$ and any 
$n \geq (R_1+R_2)^{-1}$, we have
\begin{align}
& \frac {1}{n} 
\log |{\cal X}_i^{m_i}|= \frac {m_i}{n} \log |{\cal X}_i|\leq R_i,
\notag\\
& p_{\rm e}(\phi_i^{(n)},\psi_i^{(n)}|p_{X_i}^n) \leq 
{\ExP}^{-n[E(R_i|p_{X_i})-\delta_{i,n}]}, i=1,2
\label{eqn:mainThErrB}
\end{align}
and for any eavesdropper $\A$ with $\varphi_{\A}$ satisfying
$\varphi_{\A}^{(n)} \in {\cal F}_{\A}^{(n)}(R_{\A})$, we have
\begin{align}
& \Delta^{(n)}(\varphi_1^{(n)},\varphi_2^{(n)},
       \varphi_{\A}^{(n)}|p_{X_1X_2}^n,p_{K_1K_2}^n,W^n)
\notag\\
&\leq {\ExP}^{-n[F_{\min}(R_{\A},R_1,R_2|{\pZKoKt})-\delta_{3,n}]}, 
\label{eqn:mainThSecB}
	\end{align}
where $\delta_{i,n},i=1,2,3$ are defined by
\begin{align*}
\delta_{i,n}:=&
\frac{1}{n}\log\left[{\ExP}(n+1)^{2|{\cal X}_i|}\right.
\notag\\
& \times \left. \left\{1+(n+1)^{|{\cal X}_1|}+(n+1)^{|{\cal X}_2|}\right\} 
\right],\mbox{ for }i=1,2,
\\
\delta_{3,n}:=& 
\frac{1}{n} \log \Bigl[15n(R_1+R_2)
\notag\\
& \times \Bigl\{ 1+(n+1)^{|{\cal X}|_1}+(n+1)^{|{\cal X}|_2} \Bigr\} \Bigr].
\end{align*}
Note that for $i=1,2,3$, $\delta_{i,n} \to 0$ as $n\to \infty$. 
}
\end{theorem}

This theorem is proved by a coupling of two techniques. 
One is a technique Oohama \cite{oohama:07} developed for deriving 
approximation error exponents for the intrinsic randomness problem 
in the framework of distributed random number extraction, which 
was posed by the author. This technique is used in the security 
analysis for the privacy amplification of distributed 
encrypted sources with correlated keys posed and investigated 
by Santoso and Oohama \cite{SantsoOh17EncCorK}, 
\cite{SantsoOh18PEC}. The other is a technique 
Oohama \cite{oohama2015exponent} developed 
for establishing exponential strong converse theorem 
for the one helper source coding problem. 
This technique is used in the security analysis 
for the side channel attacks to the Shannon cipher system
posed and investigated by Oohama and Santoso 
\cite{oohama17SideCh}, \cite{oohama18SideCh}.

The functions $E(R_i|p_{X_i})$ and $F(R_{\A},R_1,R_2|{\pZKoKt})$ 
take positive values 
if $(R_{\A},R_1,R_2)$ belongs to the set
\begin{align*}
&\{R_1 > H(X_1)\}\cap \{R_2 > H(X_2)\} 
                 \bigcap_{i=1,2,3}
{\cal R}_i^{\rm c}(p_{ZK_i})
\\
& 
\defeq 
{\cal R}_{\rm Sys}^{\rm (in)}
(p_{X_1X_2},{\pZKoKt}).
\end{align*}
Thus, by Theorem \ref{Th:mainth2}, under 
$$
(R_{\A},R_1,R_2) \in 
{\cal R}_{\rm Sys}^{\rm (in)}(p_{X_1X_2},{\pZKoKt}),
$$  
we have the followings: 
\begin{itemize}
\item On the reliability, 
for $i=1,2$, $p_{\rm e}(\phi_i^{(n)},\psi_i^{(n)}|p_{X_i}^n)$  
goes to zero exponentially as $n$ tends to infinity, and its 
exponent is lower bounded by the function $E(R_i|p_{X_i})$.  
\item On the security, for 
any $\varphi_{\cal A}$ satisfying 
$\varphi_{\cal A}^{(n)}\in {\cal F}_{\cal A}^{(n)}(R_{\cal A}$ $)$, 
the information leakage
$\Delta^{(n)}(
 \varphi_1^{(n)},
 \varphi_2^{(n)},
 \varphi_{\A}^{(n)}|p_{X_1X_2}^n,$ 
${\pZKoKtUn})$ on $X_1^n,X_2^n$ goes to zero exponentially 
as $n$ tends to infinity, and its exponent is lower bounded 
by the function $F_{\min}(R_{\A},R_1,R_2|p_{ZK_1K_2})$.
\item For each $i=1,2$, the code $(\phi_i^{(n)},\psi_i^{(n)})$ 
that attains the exponent function $E($$R_i|p_{X_i})$ is 
a universal code that depends only on $R_i$ not on the value 
of the distribution $p_{X_i}$. 
\end{itemize}
Define
\begin{align*}
&{\cal D}_{\rm Sys}^{\rm (in)}(p_{X_1X_1},{\pZKoKt})
:=\{(R_{\A},R_1,R_2,
\notag\\
&\quad E(R_1|p_{X_1}),
       E(R_2|p_{X_2}),
  F_{\min}(R_{\A},R_1,
           R_2|p_{K_1K_2})):
\notag\\
&\quad (R_1,R_2)\in {\cal R}_{\sf Sys}^{\rm (in)}
(p_{X_1X_2},{\pZKoKt})\}.
\end{align*}

From Theorem \ref{Th:mainth2}, we immediately obtain 
the following corollary. 
\begin{corollary}
\begin{align*}
&{\cal R}_{\rm Sys}^{\rm (in)}(p_{X_1X_1},{\pZKoKt})
 \subseteq {\cal R}_{\rm Sys}(p_{X_1X_1},{\pZKoKt}),
\\
&{\cal D}_{\rm Sys}^{\rm (in)}(p_{X_1X_1},{\pZKoKt})
 \subseteq {\cal D}_{\rm Sys}(p_{X_1X_1},{\pZKoKt}).
\end{align*}
\end{corollary}

In the remaining part of this section, we give two 
simple examples of ${\cal R}_{\rm Sys}^{\rm (in)}(p_{X_1X_1},{\pZKoKt})$. 
Those correspond two extrimal cases on the correlation of 
$(K_1,K_2,Z)$. In those two examples, we assume that 
${\cal X}_1={\cal X}_2=\{0,1\}$ and 
$p_{X_1}(1)=s_1,p_{X_2}(1)=s_2$. 
We further assume that $p_{K_1,K_2}$ 
has the binary symmetric distribution given by
$$
\ba{ll}
p_{K_1K_2}(k_1,k_2)=&(1/2)
\left[\bar{\rho} k_1 \oplus k_2 + \rho \overline{k_1 \oplus k_2}
\right] 
\\
                   &\mbox{ for }(k_1,k_2)\in \{0,1\}^2,
\ea
$$
where $\rho \in [0,0.5]$ is a parameter indicating the correlation 
level of $(K_1,K_2)$. 
\begin{Ex} We consider the case where $W=p_{Z|K_1K_2}$ is given by  
$$
\ba{ll}
W(z|k_1,k_2)=
&W(z|k_1)=
  \overline{\rho_{\cal A}}         k_1\oplus z 
      +\rho_{\cal A}\overline{k_1\oplus z}
\\
& \mbox{ for }(k_1,k_2,z) \in \{0,1\}^3.
\ea 
$$
In this case we have $K_2 \markov K_1 \markov Z$. This corresponds 
to the case where the adversary ${\cal A}$ attacks only 
node ${\rm L}_1$. Let $N_{\cal A}$ be a binary random variable 
with $p_{N_{\cal A}}(1)=\rho_{\cal A}$. We assume that   
$N_{\cal A}$ is independent of $(X_1,X_2)$ and $(K_1,K_2)$. 
Using $N_{\cal A}$, $Z$ can be written as 
$Z= K_1 \oplus N_{\cal A}.$
The inner bound for this example denoted by 
${\cal R}_{\rm Sys,{\rm ex1}}^{\rm (in)}(p_{X_1X_2},{\pZKoKt})$
is the following.
\begin{align*}
&{\cal R}_{\rm Sys, {\rm ex1}}^{\rm (in)}(p_{X_1X_2},{\pZKoKt})
=\{(R_{\cal A},R_1,R_2): 
\\
& 
\ba{rl}
0\leq R_{\cal A} & \leq 1-h(\theta),
      h(s_1)<R_1
< h( \rho_{\cal A}* \theta),
\\
      h(s_2)<R_2 & < h \left((\rho* \rho_{\cal A})* \theta \right),
\\
        R_1+R_2  & < h(\rho)+ h( \rho_{\cal A}* \theta)
\mbox{ for some }\theta \in [0,1]\},
\ea
\end{align*}
where $h(\cdot)$ denotes the binary entropy function 
and $a*b \defeq a\bar{b}+\bar{a}b$. 
\end{Ex}
\begin{Ex} We consider the case of $\rho=0.5$. In this case 
$K_1$ and $K_2$ is independent. In this case we have no information 
leakage if $R_{\cal A}=0$. We assume that $W=p_{Z|K_1K_2}$ 
is given by 
$$
\ba{ll}
W(z|k_1,k_2)=
& \overline{\rho_{\cal A}}  k_1\oplus k_2\oplus z
      +\rho_{\cal A}\overline{k_1\oplus k_2\oplus z }
\\
& \mbox{ for }(k_1,k_2,z) \in \{0,1\}^3.
\ea 
$$
Let $N_{\cal A}$ be the same random variable as the previous example. 
Using $N_{\cal A}$, $Z$ can be written as  
$
Z=K_1 \oplus K_2 \oplus N_{\cal A}.
$
The inner bound in this example denoted by 
${\cal R}_{\rm Sys,{\rm ex2}}^{\rm (in)}($ $p_{X_1X_2},{\pZKoKt})$
is the following:
\begin{align*}
&{\cal R}_{\rm Sys, {\rm ex2}}^{\rm (in)}(p_{X_1X_2},{\pZKoKt})
=\{(R_{\cal A},R_1,R_2): 
\\
& 
\ba{rl}
0\leq R_{\cal A} & \leq 1-h(\theta),
 h(s_i)<R_i 
< 1, i=1,2,
\\
   R_1+R_2  & < 1+ h(\rho_{\cal A}*\theta)
\mbox{ for some }\theta \in [0,1]\}.
\ea
\end{align*}
\end{Ex}

\begin{figure}[t]
\centering
\includegraphics[width=0.48\textwidth]{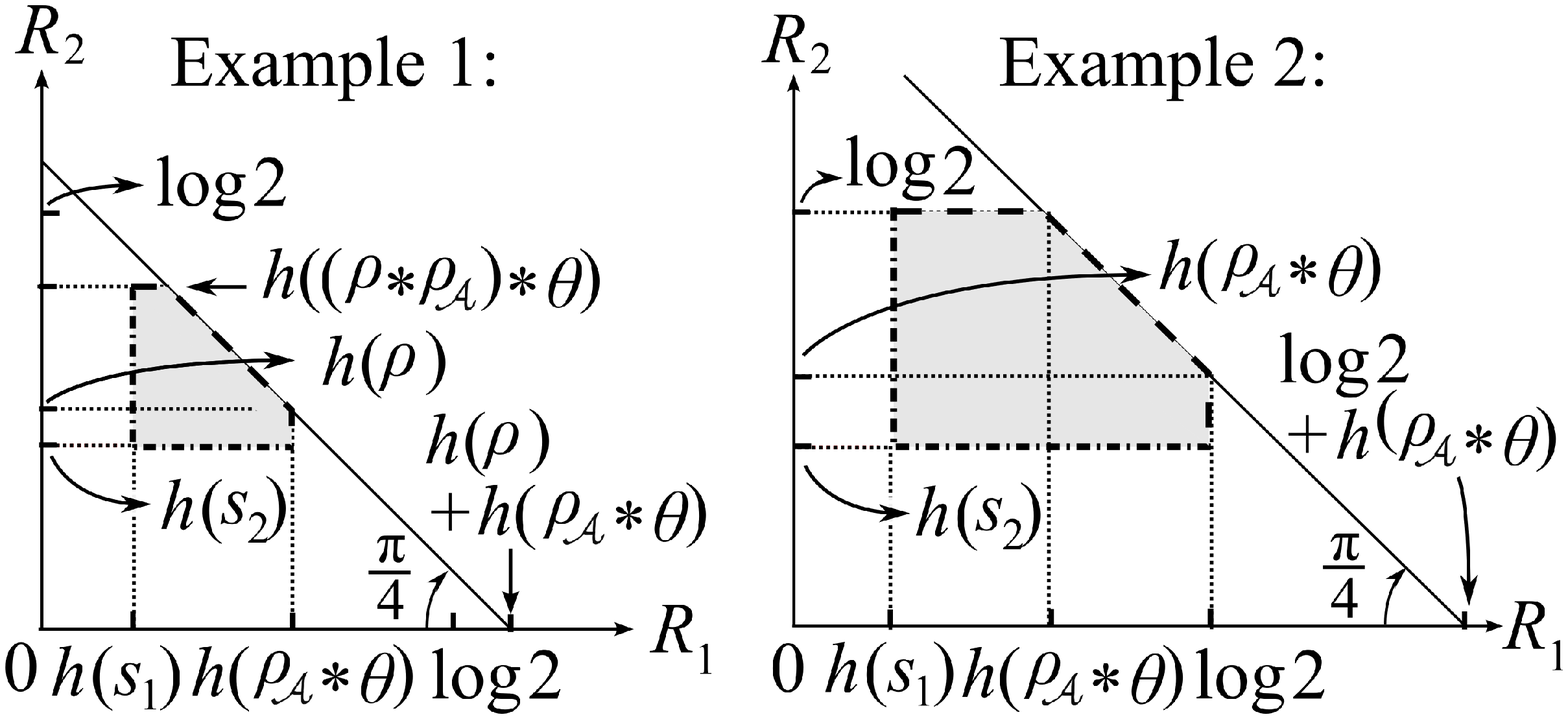}
\caption{
Shape of the regions ${\cal R}_{{\rm Sys, ex}i }^{\rm (in)}($$p_{X_1X_2},{\pZKoKt})$
$\cap \{R_{\cal A}=1-h(\theta)\}$,$i=,1,2$.
}\label{fig:RegBCCor}
\end{figure}

For the above two examples, we show the section of 
the regions 
${\cal R}_{ {\rm Sys, ex}i}^{\rm (in)}($ $p_{X_1X_2},{\pZKoKt})$
$i=,1,2$ by the plane $\{R_{\cal A}=1-h(\theta)\}$ 
is shown in Fig. \ref{fig:RegBCCor}.  

\section{Proofs of the Results}

In this section we prove Theorem \ref{Th:mainth2}. 

\subsection{Types of Sequences and Their Properties}
In this subsection we prepare basic results on the types. Those results are basic 
tools for our analysis of several bounds related to error provability of decoding 
or security.

\begin{definition}{\rm
For each $i=1,2$ and for any 
$n$-sequence ${\vcxi}=x_{i,1}x_{i,2}\cdots $ 
                     $x_{i,n} \in {\calVarX}^{n}$, 
$n(x_i|{\vcxi})$ denotes the number of $t$ such that $x_{i,t}=x_i$.  
The relative frequency 
$\left\{n(x_i|\vcxi)/n \right\}_{x_i\in {\cal X}_i}$ 
of the components of ${\vcxi}$ is called the type of ${\vcxone}$ 
denoted by $P_{\lvcxone}$.  The set that consists of all 
the types on ${\cal X}$ is denoted by ${\cal P}_{n}({\cal X})$. 
Let $\overline{X}_i$ denote an arbitrary random variable 
whose distribution 
$P_{\overline{X}_i}$ belongs to ${\cal P}_{n}({\cal X}_i)$. 
For $p_{ \overline{X}_i }\in {\cal P}_{n}( {\cal X}_i )$, set 
$$
T_{ \overline{X}_i }^n :
= \left\{ \vcxi :\, P_{\vcxi} = p_{ \overline{X}_i }\right\}.
$$
}\end{definition}

For set of types and joint types the following lemma holds. 
For the detail of the proof see Csisz\'ar 
and K\"orner \cite{csiszar-korner:11}.
\begin{lemma}\label{lem:Lem1}{\rm 
$\quad$
\begin{itemize}
\item[a)]
   $\begin{array}[t]{l} 
   |{\cal P}_{n}({\cal X}_i)|\leq(n+1)^{| {\cal X}_i |}.  
   \end{array}$
\item[b)] For $P_{\overline{X}_i}
\in {\cal P}_{n}({\cal X}_i)$,
\begin{align*}
 (n+1)^{-{|\calVarX_i|}}{\ExP}^{nH(\Dist{\overline{X}_i})}
&   \leq |T^n_{\overline{X}_i}|
\leq {\ExP}^{nH(\Dist{\overline{X}_i})}.
\end{align*}
\item[c)] For ${\vcxi} \in T^n_{\overline{X}_i}$, 
\begin{align*}
p_{X_i}^n({\vcxi})
&={\ExP}^{-n[H(\Dist{ \overline{X}_i })
          +D(p_{ \Dist{ \overline{X}_i }}||p_{X_i})]}.
\end{align*}
\end{itemize}
}
\end{lemma}

By Lemma \ref{lem:Lem1} parts b) and c), we immediately 
obtain the following lemma: 
\begin{lemma}\label{lem:Lem1.4}{\rm $\quad$
For $p_{\overline{X}_i}\in {\cal P}_{n}({\cal X}_i)$,  
\begin{align*}
& p_{X_i}^n(T_{\overline{X}_i}^n)
\leq {\ExP}^{-n D(p_{\Dist{\overline{X}_i}}||p_{X_i})}.
\end{align*}
}
\end{lemma}

\subsection{Upper Bounds on Reliablity and Security}

In this subsection we evaluate upper bounds of 
$p_{\rm e}(\phi_i^{(n)},$ $\psi_i^{(n)}|p_{X_i}^n),$ $i=1,2,$
and 
$\Delta_n (\varphi_1^{(n)},\varphi_2^{(n)},\varphi^{(n)}_{\cal A}|
p_{X_1X_2}^n, p_{ZK_1}$ ${}_{K_2Un})$. 
For $p_{\rm e}(\phi_i^{(n)}$, $\psi_i^{(n)}| p_{X_i}^n)$, 
we derive an upper bound which can be characterized with a quantity 
depending on $(\phi_i^{(n)},\psi_i^{(n)})$ and type 
$P_{\vcxi}$ of sequences $x_i^n \in {\cal X}_i^n $.
We first evaluate 
$p_{\rm e}(\phi_i^{(n)}, \psi_i^{(n)}|p_{X_i}^n), i=1,2$.   
For ${\vcxi} \in {\cal X}_i^{n}$ and 
$p_{\overline{X}}
 \in{\cal P}_{n}({\cal X}_i)$ we define the following functions.
\begin{align*}
\Xi_{{\vcxi}}(\phi_i^{(n)},\psi_i^{(n)})
&\defeq\left\{
\begin{array}{ccl}
1&\qquad&\mbox{if}\quad
         \psi_i^{(n)}\bigl(\phi_i^{(n)}({\vcxi})\bigr)
	 \neq {\vcxi},
	 \vspace{0.2cm}\\
0&\qquad&\mbox{otherwise,}
\end{array}
\right.\\
\displaystyle \Xi_{\overline{X}_i}
(\phi^{(n)},\psi^{(n)})
&\defeq \frac{1}{|T_{\overline{X}_i}^{n}|}
\sum_{{\vcxi} \in T_{\overline{X}_i}^{n}}
\Xi_{{\vcxi}}(\phi_i^{(n)},\psi_i^{(n)}).
\end{align*}
Then we have the following lemma. 
\begin{lemma} 
\label{lem:ErBound}
In the proposed system, for $i=1,2$ and for any pair of 
$(\phi_i^{(n)},$ $\psi_i^{(n)})$, we have
\begin{align}
& p_{\rm e}(\phi_i^{(n)}, \psi_i^{(n)}|p_{X_i}^n)
\notag\\
& \leq \sum_{ p_{\overline{X}_i}
     \in {\cal P}_n({\cal X}_i) }
  \Xi_{ \overline{X} }(\phi_i^{(n)},\psi_i^{(n)})
  {\ExP}^{-nD(p_{ \overline{X}_i }||p_{X_i})}.
\label{eqn:ErrBound}
\end{align}
\end{lemma}

Proof of this lemma is found in \cite{oohama17SideCh}. 
We omit the proof. 
\hfill\IEEEQED
\newcommand{\DelB}{
\begin{IEEEproof} 
We have the following chain of inequalities:
\begin{align*}
&p_{\rm e} (\phi^{(n)}, \psi^{(n)} |p_{X_1X_2}^n)
\\
&\MEq{a} \sum_{ p_{\overline{X}} 
\in {\cal P}_n({\cal X})}
\sum_{{\lvcxone}\in T^n_{ \overline{X}} }
\Xi_{\lvcxone}(\phi^{(n)},\psi^{(n)})
p_{X_1X_2}^n({\vcxone})
\\
&= \sum_{ p_ {\overline{X}} \in {\cal P}_n({\cal X})}
\frac{1}{|T^n_{\overline{X}}|}
\sum_{{\lvcxone}\in T^n_{\overline{X}}}
\Xi_{{\lvcxone} }(\phi^{(n)},\psi^{(n)})
|T^n_{\overline{X}}|p_{X}^n({\vcxone})
\\
&\MEq{b}\sum_{p_{ \overline{X}} \in {\cal P}_n({\cal X})}
\frac{1}{|T^n_{\overline{X}}|}
\sum_{\lvcxone \in T^n_{\overline{X}}}
\Xi_{ \lvcxone }(\phi^{(n)},\psi^{(n)})
p_{X}^n(T^n_{ \overline{X} })
\\
&\MEq{c}\sum_{p_{\overline{X}} 
\in {\cal P}_n({\cal X})}
\Xi_{\overline{X}}
(\phi^{(n)},\psi^{(n)})
p_{X}^n(T^n_{\overline{X}})
\\
&
\MLeq{d}\sum_{p_{\overline{X}}
\in {\cal P}_n({\cal X})}
\Xi_{\overline{X}}(\phi^{(n)},\psi^{(n)})
{\ExP}^{-nD(p_{\overline{X}}||p_{X})}.
\end{align*}
Step (a) follows from the definition of 
$\Xi_{{\lvcxone}}(\phi^{(n)},\psi^{(n)})$.
Step (b) follows from that the probabilities 
$p_{X}^n(\vcxone)$ for $ \vcxone \in T^n_{\overline{X}}$
take an identical value. 
Step (c) follows from the definition of 
$\Xi_{\overline{X}}(\phi^{(n)},\psi^{(n)})$.
Step (d) follows from lemma \ref{lem:Lem1.4}. 
\end{IEEEproof}
}

We next discuss upper bounds of 
\begin{align*}
&\Delta_{n}(\varphi_1^{(n)},\varphi_2^{(n)},
           \varphi^{(n)}_{\cal A}|p_{X_1X_2}^n,{\pZKoKtUn})
\\
&=I(\widetilde{C}_1^{m_1}
    \widetilde{C}_2^{m_2}, 
    M_{\cal A}^{(n)};X_1^{n}X_2^{n}). 
\end{align*}
On an upper bound  of 
$I(\widetilde{C}_1^{m_1}\widetilde{C}_2^{m_2}, 
 M_{\cal A}^{(n)};X_1^{n}X_2^{n})$,
we have the following lemma.
\begin{lemma}\label{lem:MIandDiv}
\begin{align}
& I(\widetilde{C}_1^{m_1} \widetilde{C}_2^{m_2}, 
    M_{\cal A}^{(n)};X_1^{n}X_2^{n}) 
\notag\\
& \leq 
\left.\left.\left.\!\!\!
D\left(p_{  K_1^{m_1}K_2^{m_2} | M_{\cal A}^{(n)} }
\right|\right|
p_{V_1^{m_1}V_2^{m_2}} \right| p_{M_{\cal A}^{(n) }}\right),
\label{eqn:oohama}
\end{align}
where $p_{V_1^{m_1}V_2^{m_2}}$ represents 
the uniform distribution over 
$\mathcal{X}_1^{m_1}\times$
$\mathcal{X}_2^{m_2}$. 
\end{lemma}

\begin{IEEEproof}
We have the following chain of inequalities: 
\begin{align*}
& I( \widetilde{C}_1^{m_1} \widetilde{C}_2^{m_2}, 
   M_{\cal A}^{(n)};{\rvcxone}{\rvcxtwo})
\MEq{a}I(\widetilde{C}_1^{m_2}
         \widetilde{C}_2^{m_2};
         {\rvcxone}{\rvcxtwo}| M_{\cal A}^{(n)})
\\
&\leq \log (|{\cal X}_1^{m_1}||{\cal X}_2^{m_2}|)
-H(\widetilde{C}_1^{m_1}
   \widetilde{C}_2^{m_2}
        |{\rvcxone}{\rvcxtwo}, M_{\cal A}^{(n)})
\\
&\MEq{b} \log (|{\cal X}_1^{m_1}||{\cal X}_2^{m_2}|)
      -H(\widetilde{K}_1^{m_1}
         \widetilde{K}_2^{m_2}|{\rvcxone}{\rvcxtwo},
                                M_{\cal A}^{(n)})
\\
& \MEq{c} \log (|{\cal X}_1^{m_1}|
                |{\cal X}_2^{m_2}|)
      -H(\widetilde{K}_1^{m_1}
         \widetilde{K}_2^{m_2}|M_{\cal A}^{(n)})
\\
&=\left.\left.\left. \!\!
    D\left(p_{K_1^{m_1}
              K_2^{m_2}|M_{{\cal A}}^{(n)}}
            \right|\right| p_{V_1^{m_1}V_2^{m_2}}
            \right|p_{M_{\cal A}^{(n)}}\right).
\end{align*}
Step (a) follows from $({\rvcxone},{\rvcxtwo})\perp M_{\cal A}^{(n)}$. 
Step (b) follows from that for $i=1,2$,
$\widetilde{C}_i^{m_i}=\widetilde{K}_i^{m_i} 
    \oplus \widetilde{X}_i^{m_i}$ 
and $\widetilde{X}_i^{m_i}=\phi_i^{(n)}({\rvcxi})$.
Step (c) follows from 
 $(\widetilde{K}_1^{m_1}, \widetilde{K}_2^{m_2},$ 
 $M_{\cal A}^{(n)})\perp ({\rvcxone},$ ${\rvcxtwo})$. 
\end{IEEEproof}

\subsection{Random Coding Arguments}

We construct a pair of affine encoders 
$
(\varphi_1^{(n)},\varphi_2^{(n)})$ 
using the random coding method. For the two decoders 
$\psi_i^{(n)},i=1,2$, we propose the minimum entropy decoder 
used in Csisz\'{a}r \cite{csiszar:82} and 
Oohama and Han \cite{oohama:94}.
 
\noindent
\underline{\it Random Construction of Affine Encoders:} \  
For each $i=1,2$, we first choose $m_i$ such that 
$$
m_i:=\left\lfloor\frac{nR_i}{\log |{\cal X}_i|}\right\rfloor,
$$
where $\lfloor a \rfloor$ stands for the integer part of $a$. 
It is obvious that for $i=1,2$, 
$$
R_i-\frac{1}{n} \leq \frac{m_i}{n}\log |{\cal X}_i| \leq R_i. 
$$
By the definition (\ref{eq:homomorphica}) of 
$\phi_i^{(n)}$, we have that for ${\vcxi} \in {\cal X}_i^n$,
\begin{align*}
& \phi_i^{(n)}({\vcxi})={\vcxi} A_i,
\end{align*}
where $A_i$ is a matrix with $n$ rows and $m_i$ columns.
By the definition (\ref{eq:homomorphic}) of 
$\varphi_i^{(n)}$, we have that for ${\vcki} \in {\cal X}_i^n$,
\begin{align*}
& \varphi_i^{(n)}({\vcki})={\vcki} A_i+b_i^{m_i},
\end{align*}
where for each $i=1,2$, $b_i^{m_i}$ is a vector with $m_i$ columns.
Entries of $A_i$ and $b_i^{m_i}$ 
are from the field of ${\cal X}_i$. Those entries 
are selected at random, 
independently of each other and 
with uniform distribution.
Randomly constructed linear encoder $\phi_i^{(n)}$ and affine encoder 
$\varphi_i^{(n)}$ have three properties shown 
in the following lemma.
\newcommand{\LemmaAffineB}{
\begin{lemma}[Properties of Linear/Affine Encoders]
\label{lem:good_set}
For each $i=1,2$, we have the following: 
\begin{itemize}
\item[a)] 
For any ${\vcxi}, {\vcvi} \in {\cal X}_i^n$ with 
         ${\vcxi} \neq {\vcvi}$, we have
	\begin{align} 
        &\Pr[\phi_i^{(n)}({\vcxi})=
        \phi_i^{(n)}({\vcvi})]=\Pr[({\vcxi} \ominus {\vcvi}) A =0^{m_i}]
        \notag \\
         &=|\mathcal{X}|^{-m_i}.
	\end{align}
\item[b)] For any ${\vcsi} \in {\cal X}_i^n$, 
         and for any $\widetilde{s}_i^{m_i}\in {\cal X}^{m_i}$, we have
	\begin{align} 
	  &\Pr[\varphi_i^{(n)}({\vcsi})=\widetilde{s}_i^{m_i}]=
          \Pr[\vcs A_i \oplus b_i^{m_i}=\widetilde{s}_i^{m_i}]
         \notag\\
         &=|\mathcal{X}_i|^{-m_i}.
	\end{align}
\item[c)] 
For any ${\vcsi},     {\vcti} \in {\cal X}_i^n$ 
   with ${\vcsi} \neq {\vcti}$, 
and for any $\widetilde{s}_i^{m_i}\in {\cal X}_i^{m_i}$, we have
	\begin{align} 
	&\Pr[\varphi_i^{(n)}({\vcsi})=
            \varphi_i^{(n)}({\vcti})=\widetilde{s}_i^{m_i}]
        \notag\\
        &= \Pr[{\vcsi} A_i \oplus b_i^{m_i}=
               {\vcti} A_i \oplus b_i^{m_i}=\widetilde{s}_i^{m_i}]
	\notag\\  
	& = |\mathcal{X}_i|^{-2m_i}.
	\end{align}
\end{itemize}
\end{lemma}
}
\LemmaAffineB

Proof of this lemma is found in \cite{oohama17SideCh}. 
We omit the proof. 
\newcommand{\ProofLemAA}{
\subsection{Proof of Lemma \ref{lem:good_set}}
\label{apd:ProofLemAA}

{\it Proof:}
Let $a_l^m$ be the $l$-th low vector of the matrix $A$.
For each $l=1,2,\cdots,n$, let $A_l^m \in{\cal X}^m$ 
be a random vector which represents 
the randomness of the choice of 
$a_l^m \in{\cal X}^m$. 
Let $B^m \in{\cal X}^m$ be a random vector which represent 
the randomness of the choice of $b^m \in{\cal X}^m$.
We first prove the part a).  
Without loss of generality we may assume 
$x_1 \neq v_1$. Under this assumption 
we have the following:
\begin{align}
& ({\vcx}\ominus \vcv ) A=0^m \Leftrightarrow 
 \sum_{l=1}^n(x_l \ominus v_l)a_l^m=0^m 
\notag\\
& \Leftrightarrow a_1^m=\sum_{l=2}^n 
\frac{v_l \ominus x_l}{x_1 \ominus v_1}a_l^m.
\label{eqn:Awff}
\end{align}
Computing $\Pr[\phi({\vcx})=\phi({\vcv})]$,
we have the following chain of equalities:
\begin{align*}
&\Pr[\phi({\vcx})=\phi({\vcv})]
=\Pr[(\vcy \ominus \vcw )A=0^m]
\\
& \MEq{a}\Pr\left[ a_1^m=\sum_{l=2}^n 
\frac{w_l\ominus y_l}{x_1 \ominus v_1}a_l^m \right]
\\
&\MEq{b}
\sum_{ \scs \left\{a_l^m \right\}_{l=2}^n 
        \atop{\scs 
        \in {\cal X}^{(n-1)m}
        }
    }   
\prod_{l=2}^n P_{A_l^m}(a_l^m)      
P_{A_1^m}\left(\sum_{l=2}^n 
\frac{w_l\ominus x_l}{y_1 \ominus v_1}a_l^m \right)
\\
& =|\mathcal{X}|^{-m}
 \sum_{ \scs \left\{a_l^m \right\}_{l=2}^n 
        \atop{
        \scs \in {\cal X}^{(n-1)m}
       }
    }   
\prod_{l=2}^n P_{A_l^m}(a_l^m)      
=|\mathcal{X}|^{-m}.
\end{align*} 
Step (a) follows from (\ref{eqn:Awff}). 
Step (b) follows from that $n$ random vecotors 
$A_l^m, l=1,2,\cdots,n$ are independent. 
We next prove the part b). We have 
the following:
\begin{align}
{\vcs} A \oplus b^{m}=\widetilde{s}^{m}
\Leftrightarrow 
b^m=\widetilde{s}^{m} \ominus 
\left\{ \sum_{l=1}^n s_la_l^m \right\}. 
\label{eqn:Awffb}
\end{align}
Computing $\Pr[{\vcs} A \oplus b^{m}=\widetilde{s}^{m}]$,
we have the following chain of equalities:
\begin{align*}
&\Pr[{\vcs} A \oplus b^{m}=\widetilde{s}^{m}]
\MEq{a}\Pr\left[b^m=\widetilde{s}^{m} \ominus 
 \left\{ \sum_{l=1}^n s_la_l^m\right\} 
\right]
\\
&\MEq{b} 
\sum_{ \scs \left\{a_l^m \right\}_{l=1}^n 
       \atop{\scs 
        \in {\cal X}^{nm}
       }
    }   
\prod_{l=1}^n P_{A_l^m}(a_l^m)      
P_{B^m}\left(
\widetilde{s}^{m} \ominus 
 \left\{ \sum_{l=1}^n s_la_l^m\right\}
\right)
\\
&=|\mathcal{X}|^{-m}
  \sum_{ \scs \left\{a_l^m \right\}_{l=1}^n 
        \atop{\scs 
        \in {\cal X}^{nm}
        }
    }   
\prod_{l=1}^n P_{A_l^m}(a_l^m)=|\mathcal{X}|^{-m}. 
\end{align*} 
Step (a) follows from (\ref{eqn:Awffb}). 
Step (b) follows from that 
$n$ random vectors $A_l^m, l=1,2, \cdots,n$ 
and $B^m$ are independent. 
We finally prove the part c). We first observe that
$
{\vcs} \neq {\vct} \Leftrightarrow 
$
is equivalent to 
$
s_i \neq t_i 
\mbox{ for some } i \in \{1,2,\cdots,n\}. 
$
Without loss of generality, we may assume that
$s_1 \neq t_1$. Under this assumption we have 
the following:
\newcommand{\Argument}{
\begin{align}
&\left.
\ba{l}
{\vcs} A \oplus b^{m}=\widetilde{s}^{m},
\\
{\vct} A \oplus b^{m}=\widetilde{t}^{m}
\ea
\right\}
\\
&\Leftrightarrow 
\left.
\ba{l}
({\vcs}\ominus {\vct})A=\widetilde{s}^m\ominus\widetilde{t}^m, 
\\
b^m=\ds \widetilde{s}^{m} \ominus 
\left\{\sum_{l=1}^n s_la_l^m \right\} 
\ea
\right\}
\notag\\
&\Leftrightarrow 
\left.
\ba{l}
\ds a_1^m=\sum_{l=2}^n 
   \frac{t_l \ominus s_l}{s_1 \ominus t_1} a_l^m
   \oplus 
\frac{\widetilde{s}^m \ominus \widetilde{t}^m}{s_1\ominus t_1},
\\
\ds b^m= \widetilde{s}^{m} \ominus 
\left\{ \sum_{l=1}^n s_l a_l^m \right\} 
\ea
\right\}
\Leftrightarrow 
\left.
\ba{l}
\ds a_1^m=\sum_{l=2}^n 
   \frac{t_l \ominus s_l}{s_1 \ominus t_1} a_l^m
   \oplus 
\frac{\widetilde{s}^m \ominus \widetilde{t}^m}{s_1\ominus t_1},
\\
\ds b^m=
\sum_{l=2}^n
\frac{t_1s_l\ominus s_1t_l}
{s_1\ominus t_1}a_l^m
\ominus \frac{t_1\widetilde{s}^m  \ominus s_1 \widetilde{t}^m }
{s_1\ominus t_1} 
\ea
\right\}.
\label{eqn:Awffc}
\end{align}
Then we have the following chain of equalities:
\begin{align*}
&\Pr[ {\vcs} A \oplus b^{m}=\widetilde{s}^{m}
\land {\vct} A \oplus b^{m}=\widetilde{t}^{m}]
\\
&\MEq{a}
\Pr\left[
a_1^m=\sum_{l=2}^n\frac{t_l \ominus s_l}{s_1 \ominus t_1} a_l^m
\oplus \frac{\widetilde{s}^m \ominus \widetilde{t}^m}{s_1\ominus t_1}
\land
b^m=
\sum_{l=2}^n
\frac{t_1s_l \ominus s_1t_l}
{s_1\ominus t_1}a_l^m
\ominus \frac{t_1\widetilde{s}^m  \ominus s_1 \widetilde{t}^m }
{s_1\ominus t_1} 
\right]
\\
&\MEq{b}
\sum_{ \scs \left\{ a_l^m \right \}_{l=2}^n 
      \atop{\scs
        \in {\cal X}^{(n-1)m}
      }
    }   
\prod_{l=2}^n P_{A_l^m}(a_l^m)      
P_{A_1^m}\left(\sum_{l=2}^n\frac{t_l \ominus s_l}{s_1 \ominus t_1} a_l^m
\oplus \frac{\widetilde{s}^m \ominus \widetilde{t}^m}{s_1\ominus t_1}
\right)
\\
&\qquad \times P_{B^m}
\left(
\sum_{l=2}^n
\frac{t_1s_l\ominus s_1t_l}
{s_1\ominus t_1}a_l^m
\ominus \frac{t_1\widetilde{s}^m  \ominus s_1 \widetilde{t}^m }
{s_1\ominus t_1} 
\right)
=|\mathcal{X}|^{-2m}
 \sum_{ \scs \left\{a_l^m \right\}_{l=2}^n 
        \atop{\scs 
        \in {\cal X}^{(n-1)m}
        }
    }   
\prod_{l=2}^n P_{A_l^m}(a_l^m)      
=|\mathcal{X}|^{-2m}.
\end{align*}
Step (a) follows from (\ref{eqn:Awffc}). 
Step (b) follows from the independent 
property on $A_l^m, l=1,2,\cdots,n$ and $B^m.$
}
\newcommand{\ArgumentB}{
\begin{align}
&
\ba{l}
{\vcs} A \oplus b^{m}=
{\vct} A \oplus b^{m}=\widetilde{s}^{m}
\ea
\notag\\
&\Leftrightarrow 
\ba{l}
({\vcs}\ominus {\vct})A=0,
b^m=\ds \widetilde{s}^{m} \ominus 
\left\{\sum_{l=1}^n s_la_l^m \right\} 
\ea
\notag\\
&\Leftrightarrow 
\ba{l}
\ds a_1^m=\sum_{l=2}^n 
   \frac{t_l \ominus s_l}{s_1 \ominus t_1} a_l^m,
\ds b^m= \widetilde{s}^{m} \ominus 
\left\{ \sum_{l=1}^n s_l a_l^m \right\} 
\ea
\notag\\
&
\Leftrightarrow 
\ba{l}
\ds a_1^m=\sum_{l=2}^n 
   \frac{t_l \ominus s_l}{s_1 \ominus t_1} a_l^m,
\ds b^m= \widetilde{s}^m \oplus 
\sum_{l=2}^n
\frac{t_1s_l\ominus s_1t_l}{s_1\ominus t_1}a_l^m.
\ea
\label{eqn:Awffc}
\end{align}
Computing 
$\Pr[{\vcs} A \oplus b^{m}={\vct} A \oplus b^{m}=\widetilde{s}^{m}]$,
we have the following chain of equalities:
\begin{align*}
&\Pr[{\vcs} A \oplus b^{m}={\vct} A \oplus b^{m}=\widetilde{s}^{m}]
\\
&\MEq{a}
\Pr\left[
a_1^m=\sum_{l=2}^n\frac{t_l \ominus s_l}{s_1 \ominus t_1} a_l^m
\right.
\\
&\qquad \left.
\land
b^m=\widetilde{s}^m \oplus 
\sum_{l=2}^n
\frac{t_1s_l \ominus s_1t_l}
{s_1\ominus t_1}a_l^m
\right]
\\
&\MEq{b}
\sum_{ \scs \left\{ a_l^m \right \}_{l=2}^n 
      \atop{\scs
        \in {\cal X}^{(n-1)m}
      }
    }   
\left[\prod_{l=2}^n P_{A_l^m}(a_l^m)\right]
P_{A_1^m}\left(\sum_{l=2}^n \frac{t_l
\ominus s_l}{s_1
\ominus t_1} a_l^m \right)
\\
&\qquad \times 
P_{B^m}
\left(
\widetilde{s}^m \oplus
\sum_{l=2}^n
\frac{t_1s_l\ominus s_1t_l}
{s_1\ominus t_1}a_l^m
\right)
\\
&=|\mathcal{X}|^{-2m}
 \sum_{ \scs \left\{a_l^m \right\}_{l=2}^n 
        \atop{\scs 
        \in {\cal X}^{(n-1)m}
        }
    }   
\prod_{l=2}^n P_{A_l^m}(a_l^m)      
=|\mathcal{X}|^{-2m}.
\end{align*}
Step (a) follows from (\ref{eqn:Awffc}). 
Step (b) follows from the independent 
property on $A_l^m, l=1,2,\cdots,n$ and $B^m.$
}
\ArgumentB
\IEEEQED
}

We next define the decoder function 
$\psi_i^{(n)}: 
{\cal X}_i^{m_i} \to {\cal X}_i^{n},i=1,2.$
To this end we define the following quantities.   
\begin{definition}{\rm
 For ${\vcxi} \in{\cal X}_i^{n}$,
we denote the entropy calculated
from the type $P_{{\vcxi}}$ by 
$H({\vcxi})$. 
In other words, for a type  
$P_{\overline{X}_i} \in {\cal P}_n({\cal X}_i)$ 
such that $P_{\overline{X}_i}=P_{{\vcxi}}$, 
we define $H({\vcxi})=H(\overline{X}_i)$.
}
\end{definition}

\noindent
\underline{\it Minimum Entropy Decoder:} \ For each $i=1,2$, and for 
$\phi_i^{(n)}(x_i^n)=\widetilde{x}_i^{m_i}$, 
we define the decoder function 
$
\psi_i^{(n)}: {\cal X}_i^{m_i}
\to {\cal X}_i^n
$ as follows:
$$
\psi_i^{(n)}(\widetilde{x}_i^{m_i})
:=\left\{\begin{array}{cl}
{\hvcxi}
   &\mbox{if } \phi_i^{(n)}({\hvcxi})=\widetilde{x}_i^{m_i},\\
   &\mbox{and }H({\hvcxi})
    <H( {\cvcxi})\\
   &\mbox{for all }{\cvcxi}\mbox{ such that }\\
   & \:\phi_i^{(n)}({\cvcxi})=\widetilde{x}_i^{m_i},\\
   & \mbox{and } 
    \:{\cvcxi} \neq {\hvcxi},
\vspace{0.2cm}\\
\mbox{arbitrary}
    & \mbox{if there is no such }{\hvcxi}
     \in{\cal X}_i^{n}.
\end{array}
\right.
$$

\noindent
\underline{\it Error Probability Bound:} \ In the following 
arguments we let expectations based on the random choice of 
the affine encoders $\varphi_i^{(n)}i=1,2$ be denoted by 
${\bf E}$$[$$\cdot]$. For, $i=1,2$, define 
$$
\Pi_{\overline{X}_i}(R_i):=
    {\ExP}^{-n[R_i-H(\overline{X}_i)]^{+}}.
$$
Then we have the following lemma.
\begin{lemma}\label{lem:LemA}
For each $i=1,2$, for any $n$ and for any
$P_{\overline{X}_i}
\in {\cal P}_{n}({\cal X}_i)$,
$$
{\bf E}\left[
\Xi_{\overline{X}_i}(\phi_i^{(n)},\psi_i^{(n)})
\right]
\leq
{\ExP}(n+1)^{|{\cal X}_i|}\Pi_{\overline{X}}(R_i).
$$
\end{lemma}

Proof of this lemma is found in \cite{oohama17SideCh}. 
We omit the proof.

\noindent
\underline{\it Estimation of Approximation Error:} \ 
Define 
\begin{align*}
&\Theta(R_1,R_2,\varphi_{\cal A}^{(n)}|{\pZKoKtUn}) 
\\
& \defeq
\sum_{\scs (a, k_1^n,k_2^n) 
\atop{\scs 
\in {\cal M}_{\cal A}^{(n)} 
\times {\cal X}_1^n 
\times {\cal X}_2^n}}
p_{M_{\cal A}^{(n)} K^n}(a,k_1^n, k_2^n ) 
\notag\\
&\quad \times \log\Bigl[1+({\rm e}^{nR_1}-1) 
 p_{K_1^n|M_{\cal A}^{(n)}}(k_1^n|a) 
\notag\\
&\qquad +({\rm e}^{nR_2}-1)p_{K_2^n|M_{\cal A}^{(n)}}(k_2^n|a)
\notag\\
& \left. \qquad +({\rm e}^{nR_1} -1)( {\rm e}^{nR_2}-1)
p_{K_1^nK_2^n|M_{\cal A}^{(n)}}(k_1^n,k_2^n|a)
\right]. 
\end{align*}

Then, we have the following lemma. 
\begin{lemma}\label{lem:LemB} \ \ For $i=1,2$ 
and for any $n,m_i$ satisfying 
$(m_i/n) \log |{\cal X}_i|$ $\leq R_i$, we have
\begin{align}
& \E \left[ D \!\!\left. \left. \left. 
\left(p_{\tilde{K}^{m_1}\tilde{K}^{m_2}|M_{{\cal A}}^{(n)}}
\right|\right|
p_{V_1^{m_1}V_2^{m_2}} \right|p_{M_{\cal A}^{(n)}}\right)\right]
\notag\\
& \leq \Theta(R_1,R_2, \varphi_{\cal A}^{(n)}|{\pZKoKtUn}).
\label{eqn:Lem2aS}
\end{align}
\end{lemma}
Proof of this lemma is given in Appendix \ref{apd:ProofLemB}. 
From the bound (\ref{eqn:Lem2aS}) 
in Lemma (\ref{lem:LemB}), we know that  
the quantity $\Theta(R_1,$ $R_2,\varphi_{\cal A}^{(n)}|{\pZKoKtUn})$
serves as an upper bound of 
the ensemble average of the conditional divergence 
$D(p_{\tilde{K}_1^{m_1}
      \tilde{K}_2^{m_2}|M_{{\cal A}}^{(n)}}$ $
||p_{V_1^{m_1}V_2^{m_2}} | p_{M_{\cal A}^{(n)}}).$

\newcommand{\ProofLemB}{
\subsection{Proof of Lemma \ref{lem:LemB}}
\label{apd:ProofLemB}


In this appendix we prove Lemma \ref{lem:LemB}. This lemma immediately 
follows from the following lemma:
\begin{lemma}\label{lem:LemBb}
\ \ For $i=1,2$ and for any $n,m_i$ satisfying 
$(m_i/n) \log |{\cal X}_i|$ $\leq R_i,$ we have  
\begin{align}
& \E \left[ D \!\! \left. \left. \left. \left(
p_{ \tilde{K}^{m_1} \tilde{K}^{m_2}| M_{\cal A}^{(n)}}
 \right|\right|
p_{ V^{m_1} V^{m_2}} \right|  p_{M_{\cal A}^{(n)}} \right) \right]
\notag\\
& \leq \sum_{\scs (a, k_1^n,k_2^n) 
\atop{\scs 
\in {\cal M}_{\cal A}^{(n)} 
\times {\cal X}_1^n 
\times {\cal X}_2^n}}
p_{M_{\cal A}^{(n)} K^n}(a,k_1^n, k_2^n )
\notag\\
&\quad \times \log \left[ 
1+(|{\cal X}_1^{m_1}|-1) 
p_{K_1^n|M_{\cal A}^{(n)}}(k_1^n|a) \right.
\notag\\
&\quad +(|{\cal X}_2^{m_2}|-1)p_{K_2^n|M_{\cal A}^{(n)}}(k_2^n|a)
\notag\\
& \left. \quad +(|{\cal X}_1^{m_1}|-1)(|{\cal X}_2^{m_2}|-1)
p_{K_1^nK_2^n|M_{\cal A}^{(n)}}(k_1^n,k_2^n|a)
\right]. 
\label{eqn:Lem2aT}
\end{align}

\end{lemma}

In fact, from $|{\cal X}_i^{m_i} |\leq {\rm e}^{nR_i}$ 
and (\ref{eqn:Lem2aT}) in Lemma \ref{lem:LemBb}, 
we have the bound (\ref{eqn:Lem2aS}) in 
Lemma \ref{lem:LemB}. In this appendix we prove Lemma \ref{lem:LemBb}. 
In the following arguments, we use the following simplified notations: 
\begin{align*}
 k_i^n , K_i^n \in {\cal X}_i^n &\Longrightarrow  k_i , K_i  \in {\cal K}_i 
\\
 \tilde{k}_i^{m_i} , \tilde{K}_i^{m_i} \in {\cal X}_i^{m_i} &\Longrightarrow 
 l_i, L_i  \in {\cal L}_i 
\\
\varphi_i^{(n)}: {\cal X}_i^n \to {\cal X}_i^{m_i} &\Longrightarrow  
\varphi_i: {\cal K}_i \to {\cal L}_i 
\\
\varphi_i^{(n)}(k_i^n)=k_i^n A_i +b_i^{m_i} 
& \Longrightarrow  \varphi_i(k_i) =k_i A_i+b_i 
\\
 V_i^{m_i} \in{\cal X}_i^{m_i} & \Longrightarrow  V_i \in {\cal L}_i
\\ 
 M_{{\cal A}}^{(n)} \in {\cal M}_{{\cal A}}^{(n)}
& \Longrightarrow  M \in {\cal M}.
\end{align*}
We define
$$
\chi_{l^{\prime},l}=
\left\{
\ba{l}
1,\mbox{ if }l^{\prime}=l, 
\\
0,\mbox{ if }l^{\prime} \neq l. 
\ea
\right.
$$
Then, the conditional distribution
of the random pair $(L_1,L_2)$ 
for given $M=a \in {\cal M}$ is 
\begin{align*}
&p_{L_1L_2|M}(l|a)
=\sum_{k \in {\cal K}} p_{K_1K_2|M} (k_1,k_2|a)
\chi_{ \varphi_1(k_1),l_1}
\chi_{ \varphi_2(k_2),l_2} 
\\
&\mbox{ for }(l_1,l_2) 
\in {\cal L}_1\times {\cal L}_2.
\end{align*}
Set 
\begin{align*}
& \Upsilon_{(\varphi_1(k_1),l_1),(\varphi_2(k_2),l_2)}
\defeq 
\chi_{ \varphi_1(k_1),l_1}
\chi_{ \varphi_2(k_2),l_2} 
\notag\\
& \quad \times 
\log \Hugebl |{\cal L}_1||{\cal L}_2| \Hugel
\sum_{\scs (k_1^{\prime},k_2^{\prime}) 
\atop{\scs \in {\cal K}_1 \times {\cal K}_2}}
p_{K_1K_2|M}(k_1^{\prime},k_2^{\prime} |a) 
\notag\\
& \qquad \times 
\chi_{ \varphi_1(k_1^{\prime}),l_1}
\chi_{ \varphi_2(k_2^{\prime}),l_2} 
\Huger \Hugebr.
\end{align*}
Then the conditional divergence  between 
$p_{L_1L_2|M}$ and $p_{V_1V_2}$ for given $M$ is given by 
\begin{align}
&\left. \left. \left. D \left(p_{L_1L_2|M}\right|\right|
p_{V_1V_2} \right | p_M \right)
=\sum_{\scs (a,k_1,k_2)
\atop{\scs \in {\cal M}\times {\cal K}_1\times{\cal K}_2}}  
\sum_{\scs (l_1,l_2) \atop{\scs \in {\cal L}_1\times{\cal L}_2}}1
\notag\\
& \quad \times 
p_{MK_1K_2}(a,k_1,k_2) 
\Upsilon_{(\varphi_1(k_1),l_1),(\varphi_2(k_2),l_2)}.
\end{align}
The quantity 
$\Upsilon_{(\varphi_1(k_1),l_1),(\varphi_2(k_2),l_2)}$
has the following form:
\begin{align}
& \Upsilon_{(\varphi_1(k_1),l_1),(\varphi_2(k_2),l_2)}
= \chi_{ \varphi_1(k_1),l_1}
  \chi_{ \varphi_2(k_2),l_2} 
\notag\\
& \times 
\log \Hugebl |{\cal L}_1||{\cal L}_2| \Hugel
p_{K_1K_2|M}(k_1,k_2 |a)
\chi_{ \varphi_1(k_1),l_1}
\chi_{ \varphi_2(k_2),l_2} 
\notag\\
& + \sum_{\scs k_2^{\prime} \in \{k_2 \}^{\rm c} }
p_{K_1K_2|M}(k_1, k_2^{\prime} |a) 
\chi_{ \varphi_1(k_1),l_1}
\chi_{ \varphi_2(k_2^{\prime}),l_2} 
\notag\\
& +\sum_{\scs k_1^{\prime} \in \{k_1 \}^{\rm c} } 
p_{K_1K_2|M}(k_1^{\prime}, k_2 |a) 
\chi_{ \varphi_1(k_1^{\prime}),l_1}
\chi_{ \varphi_2(k_2),l_2} 
\notag\\
&+ \sum_{\scs (k_1^{\prime},k_2^{\prime}) 
\atop{\scs \in \{k_1 \}^{\rm c} \times \{k_2\}^{\rm c}} }
p_{K_1K_2|M}(k_1^{\prime},k_2^{\prime} |a) 
\chi_{ \varphi_1(k_1^{\prime}),l_1}
\chi_{ \varphi_2(k_2^{\prime}),l_2} 
\Huger \Hugebr
\label{eqn:AzzxW}.
\end{align}
The above form is useful for computing   
$\E[\Upsilon_{(\varphi_1(k_1),l_1)}$ 
${}_{,(\varphi_2(k_2),l_2)}]$.

\begin{IEEEproof}[Proof of Lemma \ref{lem:LemBb}] 
Taking expectation of both  side of (\ref{eqn:AzzxW})
with respect to the random choice of the entry of the matrix $A_i$ and 
the vector $b_i$ representing the affine encoder $\varphi$, 
we have 
\begin{align}
&\left. \left. \left. \E\left[ D \left(p_{L_1L_2|M}\right|\right|
p_{V_1V_2} \right | p_M \right)\right]
=\sum_{\scs (a,k_1,k_2)
\atop{\scs \in {\cal M}
\times {\cal K}_1\times{\cal K}_2
}}  \sum_{\scs (l_1,l_2) 
\atop{\scs \in {\cal L}_1\times{\cal L}_2}
} 1
\notag\\
&\quad \times  p_{MK_1K_2}(a,k_1,k_2) 
\E \left[
\Upsilon_{(\varphi_1(k_1),l_1),(\varphi_2(k_2),l_2)}
\right].
\label{eqn:Zdxxp}
\end{align}
To compute the expectation 
$\E \left[
\Upsilon_{(\varphi_1(k_1),l_1),(\varphi_2(k_2),l_2)}
\right]$, 
we introduce an expectation operator useful for the computation.
Let  
$\E_{\varphi_1(k_1)=l_{k_1}, \varphi_2(k_2)=l_{k_2}}[\cdot]$ 
be an expectation operator based on the conditional probability measures
${\rm Pr}
\left(\cdot| \varphi_1(k_1)=l_{k_1}, 
\varphi_2(k_2)=l_{k_2}\right)$.
Using this expectation operator, the quantity 
$\E \left[
\Upsilon_{(\varphi_1(k_1),l_1),(\varphi_2(k_2),l_2)}
\right]$ can be written as 
\begin{align}
&\E \left[
\Upsilon_{(\varphi_1(k_1),l_1),(\varphi_2(k_2),l_2)}
\right]
 \notag\\
&=\sum_{\scs (l_{k_1},l_{k_2})
\atop{\scs \in {\cal L}_1\times {\cal L}_2} }  
{\rm Pr} \left( \varphi_1(k_1)=l_{k_1}, \varphi_2(k_2)=l_{k_2} \right)
\notag\\
&\quad \times \E_{\varphi_1(k_1)=l_{k_1}, 
                  \varphi_2(k_2)=l_{k_2}}
\left[\Upsilon_{(l_{k_1},l_1),(l_{k_2},l_2)}\right].
\label{eqn:SddXPP}
\end{align}
Note that 
\beq
\Upsilon_{(l_{k_1},l_1),(l_{k_2},l_2)}
=\left\{
\ba{l}
1, \mbox{ if } \varphi_1(k_1)=l_1, \varphi_2(k_2)=l_2,\\
0, \mbox{ otherwise.}
\ea
\right.
\label{eqn:SdXl}
\eeq
From (\ref{eqn:SddXPP}) and (\ref{eqn:SdXl}), we have 
\begin{align}
&\E \left[
\Upsilon_{(\varphi_1(k_1),l_1),(\varphi_2(k_2),l_2)}
\right]
={\rm Pr} \left(\varphi_1(k_1)=l_1, \varphi_2(k_2)=l_2 \right)
\notag\\
&\quad \times \E_{\varphi_1(k_1)=l_{1}, 
                  \varphi_2(k_2)=l_{2}}
\left[\Upsilon_{(l_1,l_1),(l_2,l_2)}\right]
\notag\\
&=\frac{1}{|{\cal L}_1||{\cal L}_2|}
\E_{\varphi_1(k_1)=l_{1}, 
                  \varphi_2(k_2)=l_{2}}
\left[\Upsilon_{(l_1,l_1),(l_2,l_2)}\right].
\label{eqn:ASdff}
\end{align}
Using (\ref{eqn:AzzxW}), the expectation 
$\E_{\varphi_1(k_1)=l_{1},\varphi_2(k_2)=l_{2}}
\left[\Upsilon_{(l_1,l_1),(l_2,}\right.$ $\left.{}_{l_2)}\right]$ 
can be written as 
\begin{align}
& \E_{\varphi_1(k_1)=l_{1},\varphi_2(k_2)=l_{2}}
 \left[\Upsilon_{(l_1,l_1),(l_2,l_2)}\right]
\notag\\
& =\E_{\varphi_1(k_1)=l_{1},\varphi_2(k_2)=l_{2}}\Hugebl 
  \log \Hugel |{\cal L}_1| |{\cal L}_2| \Hugecl 
   p_{K_1K_2|M}(k_1,k_2 |a) 
\notag\\
&\quad + \sum_{\scs k_2^{\prime} \in \{k_2 \}^{\rm c} }
p_{K_1K_2|M}(k_1, k_2^{\prime} |a) 
\chi_{ \varphi_2(k_2^{\prime}),l_2} 
\notag\\
&\quad +\sum_{\scs k_1^{\prime} \in \{k_1 \}^{\rm c} } 
p_{K_1K_2|M}(k_1^{\prime}, k_2 |a) 
\chi_{ \varphi_1(k_1^{\prime}),l_1}
\notag\\
&\quad + \sum_{\scs (k_1^{\prime},k_2^{\prime}) 
\atop{\scs \in \{k_1 \}^{\rm c} \times \{k_2\}^{\rm c}} }
p_{K_1K_2|M}(k_1^{\prime},k_2^{\prime} |a) 
\chi_{ \varphi_1(k_1^{\prime}),l_1}
\chi_{ \varphi_2(k_2^{\prime}),l_2} 
\Hugecr \Huger\Hugebr.
\label{eqn:AzzxWcc}
\end{align}
Applying Jensen's inequality to the right member of (\ref{eqn:AzzxWcc}), 
we obtain the following upper bound of 
$\E_{\varphi_1(k_1)=l_{1},\varphi_2(}$ ${}_{k_2)=l_{2}}
 \left[\Upsilon_{(l_1,l_1),(l_2,l_2)}\right]$
\begin{align}
& \E_{\varphi_1(k_1)=l_{1},\varphi_2(k_2)=l_{2}}
 \left[\Upsilon_{(l_1,l_1),(l_2,l_2)}\right]
\notag\\
& \leq 
\log \Hugel |{\cal L}_1| |{\cal L}_2|
     \Hugecl p_{K_1K_2|M}(k_1,k_2 |a) 
\notag\\
&\quad + \sum_{\scs k_2^{\prime} \in \{k_2 \}^{\rm c} }
p_{K_1K_2|M}(k_1, k_2^{\prime} |a) \E_2 
\notag\\
&\quad +\sum_{\scs k_1^{\prime} \in \{k_1 \}^{\rm c} } 
p_{K_1K_2|M}(k_1^{\prime}, k_2 |a) \E_1 
\notag\\
&\quad + \sum_{\scs (k_1^{\prime},k_2^{\prime}) 
\atop{\scs \in \{k_1 \}^{\rm c} \times \{k_2\}^{\rm c}} }
p_{K_1K_2|M}(k_1^{\prime}, k_2^{\prime}|a) \E_{12} 
\Hugecr \Huger, 
\label{eqn:AzzxWccd}
\end{align}
where we set 
\begin{align*}
\E_1 &\defeq 
\E_{\varphi_1(k_1)=l_1, \varphi_2(k_2)=l_2}
\left[\chi_{ \varphi_1(k_1^{\prime}),l_1}\right],
\\
\E_2 &\defeq 
\E_{\varphi_1(k_1)=l_1, \varphi_2(k_2)=l_2}
\left[\chi_{ \varphi_2(k_2^{\prime}),l_2}\right],
\\
\E_{12} &\defeq 
\E_{\varphi_1(k_1)=l_1, \varphi_2(k_2)=l_2}
\left[\chi_{ \varphi_1(k_1^{\prime}),l_1}
      \chi_{ \varphi_2(k_2^{\prime}),l_2}\right].
\end{align*} 
Computing $\E_1$, we have 
\begin{align}
\E_1&={\rm Pr}\left(\varphi_1(k_1^{\prime})=l_1
      |\varphi_1(k_1)=l_1,\varphi_2(k_2)=l_2\right)
\notag\\
   &\MEq{a}{\rm Pr}\left(\varphi_1(k_1^{\prime})=l_1
      |\varphi_1(k_1)=l_1 \right)\MEq{b}
\frac{1}{|{\cal L}_1|}.
\label{eqn:AzzxQa}
\end{align}
Step (a) follows from that the random  constructions 
of $\varphi_1$ and $\varphi_2$ are independent. 
Step (b) follows from Lemma \ref{lem:good_set} parts b) and  c). 
In a similar manner we compute $\E_2$ to obtain  
\begin{align}
\E_2&=\frac{1}{|{\cal L}_2|}.
\label{eqn:AzzxQb}
\end{align} 
We further compute $\E_{12}$ to obtain  
\begin{align}
\E_{12}
&={\rm Pr}\left(\varphi_1(k_1^{\prime})=l_1,
                \varphi_2(k_2^{\prime})=l_2
          \right.
\notag\\
&\qquad\qquad\qquad \left.
              |\varphi_1(k_1)=l_1,
               \varphi_2(k_2)=l_2\right)
\notag\\
&\MEq{a} {\rm Pr}\left(\varphi_1(k_1^{\prime})=l_1
         |\varphi_1(k_1)=l_1 \right)
\notag\\
&\quad \times 
        {\rm Pr}\left(\varphi_2(k_2^{\prime})=l_2
        |\varphi_2(k_2)=l_2 \right)
\MEq{b}\frac{1}{|{\cal L}_1||{\cal L}_2|}.
\label{eqn:AzzxQc}
\end{align}
Step (a) follows from that 
the random  constructuions of $\varphi_1$ and 
$\varphi_2$ are independent. 
Step (b) follows from Lemma \ref{lem:good_set} parts b) and  c), 
From (\ref{eqn:AzzxWccd})-(\ref{eqn:AzzxQc}), we have 
\begin{align}
&\E_{\varphi_1(k_1)=l_{1},\varphi_2(k_2)=l_{2}}
 \left[\Upsilon_{(l_1,l_1),(l_2,l_2)}\right]
\notag\\
& \leq 
\log \Hugel |{\cal L}_1||{\cal L}_2| 
     \Hugecl p_{K_1K_2|M}(k_1,k_2 |a) 
\notag\\
&\quad + \sum_{\scs k_2^{\prime} \in \{k_2 \}^{\rm c} }
p_{K_1K_2|M}(k_1, k_2^{\prime} |a)\frac{1}{|{\cal L}_2|}
\notag\\
&\quad +\sum_{\scs k_1^{\prime} \in \{k_1 \}^{\rm c} } 
p_{K_1K_2|M}(k_1^{\prime}, k_2 |a)\frac{1}{|{\cal L}_1|} 
\notag\\
&\quad + \sum_{\scs (k_1^{\prime},k_2^{\prime}) 
\atop{\scs \in \{k_1 \}^{\rm c} \times \{k_2\}^{\rm c}} }
p_{K_1K_2|M}(k_1^{\prime}, k_2^{\prime}|a) 
\frac{1}{|{\cal L}_1| |{\cal L}_2|} 
\Hugecr \Huger
\notag\\
&=\log \biggl\{1
+(|{\cal L}_1|-1)p_{K_1|M}(k_1|a) 
\notag\\
& \qquad\qquad + (|{\cal L}_2|-1)p_{K_2|M}(k_2|a) 
\notag\\
&\qquad \qquad+ (|{\cal L}_1|-1)(|{\cal L}_2|-1)
p_{K_1K_2|M}(k_1,k_2 |a)\biggr\}. 
\label{eqn:AzzxWfd}
\end{align}
From (\ref{eqn:Zdxxp}), (\ref{eqn:ASdff}), and (\ref{eqn:AzzxWfd}), 
we have the bound  (\ref{eqn:Lem2aT}) 
in Lemma \ref{lem:LemBb}.
\end{IEEEproof}
}

From Lemmas \ref{lem:MIandDiv} and \ref{lem:LemB}, we 
have the following corollary. 
\begin{corollary}\label{cor:Szz}
\begin{align*}
&{\bf E}\left[
 \Delta_{n}(\varphi_1^{(n)},
            \varphi_2^{(n)},
            \varphi_{\cal A}^{(n)}|p^n_{X_1X_2},
            {\pZKoKtUn})\right]
\notag\\
& \leq  \Theta(R_1,R_2,\varphi_{\cal A}^{(n)}|{\pZKoKtUn}).
\end{align*}
\end{corollary}

\noindent
\underline{\it Existence of Universal Code 
$
\{(\varphi_i^{(n)},\psi_i^{(n)})\}_{i=1,2}
$:} 

From Lemma \ref{lem:LemA} and Corollary \ref{cor:Szz}, 
we have the following lemma stating an existence 
of universal code 
$\{(\varphi_i^{(n)},
       \psi_i^{(n)})\}_{i=1,2}$. 
\begin{lemma}\label{lem:UnivCodeBounda} There exists at 
least one deterministic code  
$\{(\varphi_i^{(n)},\psi_i^{(n)})\}_{i=1,2}$ satisfying 
$(m_i/n)\log |{\cal X}_i|\leq R_i,i=1,2$, such that 
for $i=1,2$ and for any 
$p_{\overline{X}_i}$ $\in {\cal P}_n({\cal X}_i)$,
\begin{align*}
&\Xi_{\overline{X}_i}(\phi_i^{(n)},\psi_i^{(n)})
\leq {\ExP}(n+1)^{|{\cal X}_i|}
\\
&\times \{1+(n+1)^{|{\cal X}_1|}
           +(n+1)^{|{\cal X}_2|}\}\Pi_{\overline{X}_i}(R_i).
\end{align*}
Furthermore, for any $\varphi_{\cal A}^{(n)} 
\in {\cal F}_{\cal A}^{(n)}(R_{\cal A})$, we have  
\begin{align*}
&\Delta_{n}(\varphi_1^{(n)},
            \varphi_2^{(n)},
     \varphi_{\cal A}^{(n)}|p^n_{X_1X_2}, {\pZKoKtUn})
\\
& \leq \{1+(n+1)^{|{\cal X}_1|}
          +(n+1)^{|{\cal X}_2|}\}
\Theta(R_1,R_2, \varphi_{\cal A}^{(n)}|{\pZKoKtUn}).
\end{align*}
\end{lemma}
\begin{IEEEproof}
We have the following chain of inequalities:
\begin{align}
&{\bf E}
\left[
\frac{\Delta_n(\varphi_1^{(n)},
               \varphi_2^{(n)},
               \varphi_{\cal A}^{(n)}|
 p_{X_1X_2}^n,{\pZKoKtUn})}
{\Theta(R_1,R_2,\varphi_{\cal A}^{(n)}|{\pZKoKtUn})}
\right.
\notag\\
&\quad 
+\sum_{i=1,2}
\sum_{
p_{\overline{X}_i} 
\in {\cal P}_n({\cal X}_i)
}
\frac{\Xi_{\overline{X}_i}( \phi_i^{(n)}, \psi_i^{(n)})}
{{\ExP}(n+1)^{|{\cal X}_i|}\Pi_{\overline{X}_i}(R_i)}
\Hugebr
\notag\\
&=\frac{{\bf E}\left[\Delta_n(\varphi_1^{(n)},\varphi_2^{(n)},
\varphi_{\cal A}^{(n)}|p_{X_1X_2}^n,{\pZKoKtUn})\right]}
   {\Theta(R_1,R_2,\varphi_{\cal A}^{(n)}|{\pZKoKtUn})}
\notag\\
&\quad + 
\sum_{i=1,2}
\sum_{
p_{\overline{X}_i} 
\in {\cal P}_n({\cal X}_i)
}
\frac{{\bf E}\left[
\Xi_{\overline{X}_i}(\phi_i^{(n)},\psi_i^{(n)})
\right]}
{{\ExP}(n+1)^{|{\cal X}_i|}\Pi_{\overline{X}_i}(R_i)}
\notag\\
&\MLeq{a} 
1+\sum_{i=1,2}
\sum_{p_{\overline{X}_i}\in {\cal P}_n({\cal X}_i)}1
\MLeq{b} 1 + \sum_{i=1,2}(n+1)^{|{\cal X}_i|}.
\notag
\end{align}
Step (a) follows from Lemma \ref{lem:LemA} and 
Corollary \ref{cor:Szz}. Step (b) follows 
from Lemma \ref{lem:Lem1} part a).
Hence there exists at least one deterministic code
$\{(\varphi_i^{(n)},\psi_i^{(n)})\}_{i=1,2}$ such that
\begin{align}
&\frac{\Delta_n(\varphi_1^{(n)},
                \varphi_2^{(n)},
                \varphi_{\cal A}^{(n)}
|p_{X_1X_2}^n,{\pZKoKtUn})}
{\Theta(R_1,R_2,\varphi_{\cal A}^{(n)}|{\pZKoKtUn})}
+\sum_{i=1,2}\sum_{
p_{\overline{X}_i} 
\in {\cal P}_n({\cal X}_i)
}1
\notag\\
&\times 
\frac{\Xi_{\overline{X}_i}(\phi_i^{(n)},\psi_i^{(n)})}
{{\ExP}(n+1)^{|{\cal X}_i|}\Pi_{\overline{X}_i}(R_i)}
\leq 1+ \sum_{i=1,2}(n+1)^{|{\cal X}_i|},
\notag
\end{align}
from which we have that for $i=1,2$ and for any 
$p_{\overline{X}_i}\in {\cal P}_n({\cal X}_i)$, 
\begin{align*}
&\frac{\Xi_{\overline{X}_i}(\phi^{(n)}_i,\psi^{(n)}_i)}
{{\ExP}(n+1)^{|{\cal X}_i|}\Pi_{\overline{X}_i}(R_i)}
\leq 1+\sum_{i=1,2} (n+1)^{|{\cal X}_i|}.
\end{align*}
Furthermore, we have that for any 
$\varphi_{\cal A}^{(n)}
\in {\cal F}_{\cal A}^{(n)}(R_{\cal A})$,
\begin{align*}
&\frac{\Delta_n(\varphi_1^{(n)},\varphi_2^{(n)},
       \varphi_{\cal A}^{(n)}|p_{X_1X_2}^n,{\pZKoKtUn})}
{{\Theta(R_1,R_2,\varphi_{\cal A}^{(n)}|{\pZKoKtUn})}}
\leq 1+ \sum_{i=1,2} (n+1)^{|{\cal X}_i|},
\end{align*}
completing the proof.
\end{IEEEproof}

%

\begin{proposition}\label{pro:UnivCodeBound} 
For any $R_{\cal A}, R_1,R_2>0$, and 
any $p_{ZK_1K_2}$, there exist two sequences of mappings 
$\{(\varphi_i^{(n)}, \psi_i^{(n)}) \}_{n=1}^{\infty},i=1,2$
such that for $i=1,2$ and for any 
$p_{X_i}\in {\cal P}({\cal X}_i)$, we have 
\begin{align}
& \frac {1}{n} 
\log |{\cal X}_i^{m_i}|= \frac {m_i}{n} \log |{\cal X}_i|\leq R_i,
\notag\\
&  p_{\rm e}(\phi_i^{(n)},\psi_i^{(n)}|p_{X_i}^n) 
   \leq  {\ExP}(n+1)^{2|{\cal X}_i|}
\notag\\
&\quad \times \{1+ (n+1)^{|{\cal X}_1|}
                  +(n+1)^{|{\cal X}_2|}\}{\ExP}^{-nE(R_i|p_{X_i})} 
\label{eqn:mainThErrBa}
\end{align}
and for any eavesdropper $\A$ with $\varphi_{\A}$ satisfying
$\varphi_{\A}^{(n)} \in {\cal F}_{\A}^{(n)}(R_{\A})$, we have
\begin{align}
& \Delta^{(n)}(\varphi_1^{(n)}, \varphi_2^{(n)},
               \varphi_{\A}^{(n)}|p_{X_1X_2}^n,{\pZKoKtUn})
\notag\\
&\leq \{1+(n+1)^{|{\cal X}_1|}+(n+1)^{|{\cal X}_2|} \}
\notag\\
&\quad \times \Theta(R_1,R_2,\varphi_{\cal A}^{(n)}|{\pZKoKtUn}).
\label{eqn:mainThSecBa}
\end{align}
\end{proposition}

\begin{IEEEproof}
By Lemma \ref{lem:UnivCodeBounda}, there exists 
$(\varphi_i^{(n)},$ $\psi_i^{(n)}),i=1,2,$ satisfying 
$(m_i/n)\log |{\cal X}_i|\leq R_i$, such that 
for $i=1,2$ and for any $p_{\overline{X}_i}$ 
$\in {\cal P}_n({\cal X}_i)$,
\begin{align}
&\Xi_{\overline{X}_i}(\phi_i^{(n)},\psi_i^{(n)})
 \leq {\ExP}(n+1)^{|{\cal X}_i|}
\notag\\
&\quad \times \{1+(n+1)^{|{\cal X}_1|}+(n+1)^{|{\cal X}_2|}\}
\Pi_{\overline{X}}(R_i).
\label{eqn:aaSSS}
\end{align}
Furthermore for any $\varphi_{\cal A}^{(n)}
\in {\cal F}_{\cal A}^{(n)}(R_{\cal A})$,
\begin{align}
&\Delta_n(\varphi_1^{(n)},\varphi_2^{(n)},\varphi_{\cal A}^{(n)}
|p_{X_1X_2}^n,{\pZKoKtUn})
\notag\\
&\leq \{1+(n+1)^{|{\cal X}_1|}
         +(n+1)^{|{\cal X}_2|} \}
\notag\\
&\quad \times \Theta(R_1,R_2,\varphi_{\cal A}^{(n)}|{\pZKoKtUn}).
\label{eqn:abSSSz}
\end{align}
The bound (\ref{eqn:mainThSecBa}) in 
Proposition \ref{pro:UnivCodeBound}
has already been proved in (\ref{eqn:abSSSz}). 
Hence it suffices to prove the bound (\ref{eqn:mainThErrBa}) 
in Proposition \ref{pro:UnivCodeBound} to complete the proof.
On an upper bound of 
$p_{\rm e}(\phi_i^{(n)},
           \psi_i^{(n)}|p_{X_i}^n),i=1,2$,  
we have the following chain of inequalities:  
\begin{align*}
& p_{\rm e}(\phi_i^{(n)},\psi_i^{(n)}|p_{X_i}^n)  
 \MLeq{a} {\ExP}(n+1)^{|{\cal X}_i|}
\\
&\qquad\qquad\times \{1+(n+1)^{|{\cal X}_1|}
                       +(n+1)^{|{\cal X}_2|}\}
\\
&\qquad\qquad \times \sum_{p_{\overline{X}_i}
   \in {\cal P}_n({\cal X}_i) }
     \Pi_{\overline{X}_i}(R_i)
   {\ExP}^{-nD(p_{\overline{X}_i}||p_{X_i})}
\\
&\leq {\ExP}(n+1)^{|{\cal X}_i|}\{ (n+1)^{|{\cal X}_i|}+1\}
|{\cal P}_n({\cal X}_i)|{\ExP}^{-nE(R_i|p_{X_i})} 
\\
&\MLeq{b} {\ExP}(n+1)^{2|{\cal X}_i|}
          \{1+(n+1)^{|{\cal X}_1|}
             +(n+1)^{|{\cal X}_2|}\} 
\\
&\quad \times {\ExP}^{-nE( R_i| p_{X_i})}.
\end{align*}
Step (a) follows from Lemma \ref{lem:ErBound}
and (\ref{eqn:aaSSS}).
Step (b) follows from Lemma \ref{lem:Lem1} part a).
\end{IEEEproof}

\subsection{Explicit Upper Bound of 
$\Theta(R_1,R_2,\varphi_{\cal A}^{(n)}|{\pZKoKtUn})$} 

In this subsection we derive an explicit upper bound of 
$\Theta(R_1,R_2,\varphi_{\cal A}^{(n)}|{\pZKoKtUn})$ 
which holds for any eavesdropper 
$\A$ with $\varphi_{\A}$ satisfying
$\varphi_{\A}^{(n)} \in {\cal F}_{\A}^{(n)}(R_{\A})$. 
Define
\begin{align*}
& \wp_0 \defeq p_{M_{\cal A}^{(n)}Z^nK_1^nK_2^n}\Biggl\{
\notag\\
&\quad
\ba{rl}
R_1 &\geq \ds \frac{1}{n} 
\log \frac{1}{p_{K_1^n|M_{\cal A}^{(n)}}(K_1^n|M_{\cal A}^{(n)})}
    -\eta\\
&\mbox{or}\vspace*{-2mm}\\
 R_2 &\geq \ds \frac{1}{n} 
\log \frac{1}{p_{K_2^n|M_{\cal A}^{(n)}}(K_2^n|M_{\cal A}^{(n)})}
     -\eta_2\\
&\mbox{or}\vspace*{-2mm}\\
 R_1+R_2&\geq \ds \frac{1}{n} 
\log \frac{1}
{p_{K_1^nK_2^n|M_{\cal A}^{(n)}}(K_1^n,K_2^n|M_{\cal A}^{(n)})}
     -\eta_3 \Biggr\}.
\ea
\end{align*}
For $i=1,2$, define
\begin{align*}
&\wp_i \defeq p_{ M_{\cal A}^{(n)} Z^n K_i^n}\Biggl\{
\notag\\
& R_i \geq \left.
 \frac{1}{n}\log\frac{1}{p_{K_i^n|M_{\cal A}^{(n)}}(K_i^n|M_{\cal A}^{(n)})}
-\eta_i
\right\}.
\end{align*}
Furthermore, define
\begin{align*}
& \wp_3 \defeq p_{ M_{\cal A}^{(n)} Z^nK_1^nK_2^n}\Biggl\{
\nonumber\\
& R_1+R_2 \geq \left.
\frac{1}{n}\log
\frac{1}{p_{K_1^nK_2^n|M_{\cal A}^{(n)}}(K_1^n,K_2^n|M_{\cal A}^{(n)})}
-\eta_3
\right\}.
\end{align*}
By definition it is obvious that
\begin{align}
 \wp_0 &\leq \sum_{i=1}^3\wp_i.
\label{eqn:azcaPz}
\end{align}
We have the following lemma. 
\begin{lemma}\label{lem:ThetaBound}
For any $\eta_i>0,i=1,2,3$ and for any eavesdropper $\A$ with 
$\varphi_{\A}$ satisfying
$\varphi_{\A}^{(n)} \in {\cal F}_{\A}^{(n)}(R_{\A})$, we have 
the following: 
\begin{align}
& \Theta(R_1,R_2, \varphi_{\cal A}^{(n)} |{\pZKoKtUn})
\notag\\
& \leq n(R_1+R_2)\wp_0 +\sum_{i=1}^3{\rm e}^{-n\eta_i} 
\label{eqn:azca}
\\
& \leq n(R_1+R_2)\left[\sum_{i=1}^3 \wp_i \right] 
+\sum_{i=1}^3{\rm e}^{-n\eta_i} 
\label{eqn:azcaB}
\end{align}
Specifically, if $n \geq [R_1+R_2]^{-1}$, we have
\beqa
& &(n[R_1+R_2])^{-1}
\Theta(R_1,R_2, \varphi_{\cal A}^{(n)} |{\pZKoKtUn})
\nonumber\\
& &\leq \sum_{i=1}^3(\wp_i+{\rm e}^{-n\eta_i}). 
\label{eqn:Zddaa} 
\eeqa
\end{lemma}

\begin{IEEEproof}
By (\ref{eqn:azcaPz}), it sufficies to show 
(\ref{eqn:azca}) to prove Lemma \ref{lem:ThetaBound}.
We set 
\begin{align*}
&A_{R_1,R_2}(K_1^n,K_2^n |M_{{\cal A}}^{(n)})
\\
& \defeq 
\:({\rm e}^{nR_1}-1)p_{K_1^n|M_{\cal A}^{(n)}} 
            (K_1^n| M_{{\cal A}}^{(n)})
\notag\\
&\quad +({\rm e}^{nR_2}-1) p_{K_2^n|M_{{\cal A}}^{(n)}}
           (K_2^n|M _{{\cal A}}^{(n)} )
\notag\\
&\quad  +({\rm e}^{nR_1} -1)({\rm e}^{nR_2 }-1)
p_{K_1^nK_2^n|M_{{\cal A}}^{(n)}}
                     (K_1^n,K_2^n| M_{{\cal A}}^{(n)}).
\end{align*}
Then we have 
\begin{align}
&\Theta(R_1,R_2,\varphi_{\cal A}^{(n)}|{\pZKoKtUn}) 
\notag\\
&= {\rm E} \left[
\log \Bigl\{1+A_{R_1,R_2}(K_1^n,K_2^n |M_{{\cal A}}^{(n)})
     \Bigr\}\right]. 
\label{eqn:AdttA}
\end{align}
\newcommand{\PartS}{
\log\Bigl\{1+({\rm e}^{nR_1}-1)p_{K_1^n|M_{\cal A}^{(n)}} 
            (K_1^n| M_{{\cal A}}^{(n)})
\notag\\
&\quad      +({\rm e}^{nR_2}-1)p_{K_2^n|M_{\cal A}^{(n)}}
           (K_2^n|M _{{\cal A}}^{(n)} )
\notag\\
&\quad  +({\rm e}^{nR_1}-1)({\rm e}^{nR_2 }-1)
\notag\\
&\qquad \times   p_{K_1^nK_2^n|M_{{\cal A}}^{(n)}}
                      (K_1^n,K_2^n| M_{{\cal A}}^{(n)})
\Bigr\}
}
We further observe the following: 
\begin{align}
&\left\{
\ba{rl}
 R_1 &< \ds \frac{1}{n} 
\log \frac{1}{p_{K_1K_2^n|M_{\cal A}^{(n)}}(K^n|M_{\cal A}^{(n)})}
     -\eta_1 \vspace*{1mm}\\
 R_2 &< \ds \frac{1}{n} 
\log \frac{1}{p_{K_1K_2^n|M_{\cal A}^{(n)}}(K^n|M_{\cal A}^{(n)})}
     -\eta_2\vspace*{1mm}\\
 R_1+R_2&< \ds \frac{1}{n} 
\log \frac{1}{p_{K_1K_2^n|M_{\cal A}^{(n)}}(K^n|M_{\cal A}^{(n)})}
     -\eta_3
\ea
\right.
\notag\\
&\Rightarrow
A_{R_1,R_2}(K_1^n,K_2^n|M_{\cal A}^{(n)})
<\sum_{i=1}^3{\rm e}^{-n\eta_i}
\notag \\
&\MRarrow{a}
\log \left\{1+A_{R_1,R_2}
(K_1^n,K_2^n|M_{\cal A}^{(n)})\right\}
\leq \sum_{i=1}^3{\rm e}^{-n\eta_i}.
\label{eqn:AdttB}
\end{align}
Step (a) follows from $\log (1+a)\leq a$.
We also note that
\begin{align}
& \log \Bigl\{1+({\rm e}^{nR_1}-1)p_{K_1^n|M_{\cal A}^{(n)}} 
              (K_1^n| M_{{\cal A}}^{(n)})
\notag\\
&\quad      +({\rm e}^{nR_2}-1)p_{K_2^n|M_{\cal A}^{(n)}}
             (K_2^n|M _{{\cal A}}^{(n)})
\notag\\
&\quad  +({\rm e}^{nR_1}-1)({\rm e}^{nR_2 }-1)
\notag\\
&\qquad \times  p_{K_1^nK_2^n|M_{\cal A}^{(n)}}
                  (K_1^n,K_2^n|M_{\cal A}^{(n)})
\Bigr\}
\notag\\
& \leq \log [{\rm e}^{nR_1} {\rm e}^{nR_2}]=n(R_1+R_2).
\label{eqn:AdttC}
\end{align}
From (\ref{eqn:AdttA}),
     (\ref{eqn:AdttB}),
     (\ref{eqn:AdttC}), we have
the bound (\ref{eqn:azca}). 
\end{IEEEproof}

On upper bound of 
$\wp_i,i=1,2,3$, we have 
the following lemma:
\begin{lemma}
\label{lem:Ohzzz}
For any $\eta>0$ and for any eavesdropper 
$\A$ with $\varphi_{\A}$ satisfying
$\varphi_{\A}^{(n)} \in {\cal F}_{\A}^{(n)}(R_{\A})$,
we have that for each $i=1,2$, 
we have $\wp_i \leq \tilde{\wp}_i$, where 
\begin{align}
& \tilde{\wp}_i \defeq p_{M_{\cal A}^{(n)}Z^nK_i^n}\Biggl\{
\notag\\
0& \geq 
\frac{1}{n}\log \frac{ 
\hat{q}_{i, M_{\cal A}^{(n)} Z^n K_i^n }(M_{\cal A}^{(n)},Z^n,K_i^n)}
{p_{M_{\cal A}^{(n)}Z^nK^n}(M_{\cal A}^{(n)},Z^n,K_i^n)}-\eta_i,
\label{eqn:asppa}\\
0& \geq \frac{1}{n}\log \frac{ Q_{i,{Z}^n} (Z^n) }{p_{Z^n}(Z^n)}-\eta_i,
\label{eqn:asppb}\\
R_{\cal A}&\geq \ds \frac{1}{n}\log
\frac{Q_{i,Z^n|M_{\cal A}^{(n)}}(Z^n|M_{\cal A}^{(n)})}
     {p_{Z^n}(Z^n)}-\eta_i,
\label{eqn:asppc}\\
R_i&\geq \ds \frac{1}{n}\log 
\frac{1}{Q_{i, K_i^n| M_{\cal A}}^{(n)}(K_i^n|M_{\cal A}^{(n)})}
-\eta_i
\Biggr\}
+3{\rm e}^{-n\eta_i}. 
\label{eqn:azsad}
\end{align}
and that for $i=3$, we have $\wp_3\leq \tilde{\wp}_3$, where 
\begin{align}
& \tilde{\wp}_3 \defeq p_{M_{\cal A}^{(n)}Z^nK_1^nK_2^n}\Biggl\{
\notag\\
0 \geq  \frac{1}{n} & \log
\frac{\hat{q}_{3, M_{\cal A}^{(n)}Z^nK_1^nK_2^n}
(M_{\cal A}^{(n)},Z^n,K_1^n,K_2^n)}
{p_{M_{\cal A}^{(n)}Z^nK_1^nK_2^n}(M_{\cal A}^{(n)},Z^n,K_1^nK_2^n)}-\eta_3,
\label{eqn:asppaG}\\
0 \geq \frac{1}{n} & \log \frac{ Q_{3,{Z}^n}(Z^n)}{p_{Z^n}(Z^n)}-\eta_3,
\label{eqn:asppbG}\\
R_{\cal A}&\geq \ds \frac{1}{n}\log
\frac{\tilde{Q}_{3, Z^n|M_{\cal A}^{(n)}}(Z^n|M_{\cal A}^{(n)})}
{p_{Z^n}(Z^n)}-\eta_3,
\label{eqn:asppcG}\\
R_1+R_2&\geq \ds \frac{1}{n}\log 
\frac{1}{p_{K_1^nK_2^n|M_{\cal A}^{(n)}}(K_1^n,K_2^n|M_{\cal A}^{(n)})}
-\eta_3
\Biggr\}
\notag\\ 
&\quad +3{\rm e}^{-n\eta_3}. \label{eqn:azsadG}
\end{align}
The probability distributions appearing 
in the three inequalities 
(\ref{eqn:asppa}), 
(\ref{eqn:asppb}), and 
(\ref{eqn:asppc}) 
in the right members of (\ref{eqn:azsad}) have 
a property that we can select them arbitrary. 
In (\ref{eqn:asppa}), we can choose any probability 
distribution $\hat{q}_{i, M_{\cal A}^{(n)}Z^nK_i^n}$ on 
${\cal M}_{\cal A}^{(n)}$ $\times{\cal Z}^n$ $\times{\cal X}_i^n$. 
In (\ref{eqn:asppb}), we can choose any 
distribution $Q_{i, Z^n}$ on ${\cal Z}^n$. 
In (\ref{eqn:asppc}), we can choose any 
stochastic matrix 
$\tilde{Q}_{i,Z^n|M_{\cal A}^{(n)}}$: 
${\cal M}_{\cal A}^{(n)}$ $\to {\cal Z}^n$. 
The probability distributions appearing in the three inequalities 
(\ref{eqn:asppaG}), (\ref{eqn:asppbG}), and (\ref{eqn:asppcG}) 
in the right members of (\ref{eqn:azsadG}) have 
a property that we can select them arbitrary. 
In (\ref{eqn:asppaG}), we can choose any probability 
distribution 
$\hat{q}_{3, M_{\cal A}^{(n)} Z^n K_1^n K_2^n}$ 
on ${\cal M}_{\cal A}^{(n)}$ 
$\times{\cal Z}^n$ 
$\times{\cal X}_1^n$ 
$\times{\cal X}_2^n$. 
In (\ref{eqn:asppbG}), we can choose any 
distribution ${Q}_{3,Z^n}$ on ${\cal Z}^n$. 
In (\ref{eqn:asppcG}), we can choose any 
stochastic matrix 
$\tilde{Q}_{3, Z^n|M_{\cal A}^{(n)}}$: 
${\cal M}_{\cal A}^{(n)}$ $\to {\cal Z}^n$. 
\end{lemma}

The above lemma is the same as Lemma 10 in the previous 
work \cite{oohama17SideCh}. Since the proof of the lemma 
is in \cite{oohama17SideCh}, 
we omit the proof of Lemma \ref{lem:Ohzzz} in the present paper.
We have the following proposition.
\begin{proposition}\label{pro:ThetaExpUpper} 
For any 
$\varphi_{\cal A}^{(n)} 
\in {\cal F}_{\cal A}^{(n)}(R_{\cal A})$ and any 
$n \geq [R_1+R_2]^{-1}$, we have
\begin{align}
&(n[R_1+R_2])^{-1}\Theta(R_1,R_2,\varphi_{\cal A}^{(n)}|p^n_{ZK_1K_2})
\notag\\
& \leq 15{\ExP}^{-nF_{\min}(R_{\cal A},R_1,R_2|p_{ZK_1K_2})}.
\label{eqn:SSssP}
\end{align}
\end{proposition}
{\it Proof:} By Lemmas \ref{lem:ThetaBound} and \ref{lem:Ohzzz}, 
we have for any 
\begin{align}
& (n[R_1+R_2])^{-1}\Theta(R_1,R_2,\varphi_{\cal A}^{(n)}|p^n_{ZK_1K_2})
\notag\\
& \leq \sum_{i=1}^{3}
(\tilde{\wp}_i+{\rm e}^{-n\eta_i}).
\label{eqn:SSssPa}
\end{align}
The quantity $\tilde{\wp}_i+{\rm e}^{-n\eta_i},i=1,2,3$.
is the same as the upper bound on 
the correct probability of decoding for one helper 
source coding problem in Lemma 1 in 
Oohama \cite{oohama2015exponent}(extended version). 
In a manner similar to the derivation of the exponential upper 
bound of the correct probability of decoding for one helper 
source coding problem, we can prove that 
for any $\varphi_{\cal A}^{(n)}\in {\cal F}_{\cal A}^{(n)}(R_{\cal A})$ 
there exist $\eta_i^*,i=1,2,3$ such that for $i=1,2,3$, 
we have  
\begin{align}
\tilde{\wp}_i+ {\rm e}^{-n\eta_i^*}
\leq 5{\ExP}^{-nF(R_{\cal A},R_i|p_{ZK_i})}.
\label{eqn:SSssPb} 
\end{align}
From (\ref{eqn:SSssPa}) and (\ref{eqn:SSssPb}), we have
that for any 
$\varphi_{\cal A}^{(n)} 
\in {\cal F}_{\cal A}^{(n)}(R_{\cal A})$ and any 
$n \geq [R_1+R_2]^{-1}$, 
\begin{align*}
&(n[R_1+R_2])^{-1}
\Theta(R_1,R_2,\varphi_{\cal A}^{(n)}|p^n_{ZK_1K_2})
\notag\\
& \leq 5 \sum_{i=1}^3{\ExP}^{-nF(R_{\cal A},R_i|p_{ZK_i})}
\leq 15{\ExP}^{-nF_{\min}(R_{\cal A},R_1,R_2|p_{ZK_1K_2})},
\end{align*}
completing the proof. 
\hfill\IEEEQED

\appendix


\ProofLemB

\bibliographystyle{IEEEtran}

\bibliography{RefEdByOh}

\end{document}